\documentclass[acmlarge, nonacm=true]{acmart}

\usepackage{fnpct}
\usepackage{multirow}
\usepackage{algpseudocode}
\usepackage{algorithm}
\usepackage{graphicx}
\usepackage{subcaption}
\usepackage{array}
\usepackage[export]{adjustbox}
\usepackage{flushend}
\usepackage{hyperref}
\usepackage{listings}
\usepackage{xcolor}
\usepackage[utf8]{inputenc}

\lstdefinelanguage{JavaScript}{
  keywords={typeof, new, true, false, catch, function, return, null, catch, switch, var, if, in, while, do, else, case, break},
  keywordstyle=\color{blue}\bfseries,
  ndkeywords={class, export, boolean, throw, implements, import, this},
  ndkeywordstyle=\color{darkgray}\bfseries,
  identifierstyle=\color{black},
  sensitive=false,
  comment=[l]{//},
  morecomment=[s]{/*}{*/},
  commentstyle=\color{purple}\ttfamily,
  stringstyle=\color{red}\ttfamily,
  morestring=[b]',
  morestring=[b]"
}

\definecolor{codegreen}{rgb}{0,0.6,0}
\definecolor{codegray}{rgb}{0.5,0.5,0.5}
\definecolor{codepurple}{rgb}{0.58,0,0.82}
\definecolor{backcolour}{rgb}{0.95,0.95,0.92}

\lstdefinestyle{mystyle}{
    backgroundcolor=\color{backcolour},   
    commentstyle=\color{codegreen},
    keywordstyle=\color{magenta},
    numberstyle=\tiny\color{codegray},
    stringstyle=\color{codepurple},
    basicstyle=\ttfamily\footnotesize,
    breakatwhitespace=false,         
    breaklines=true,                 
    captionpos=b,                    
    keepspaces=true,                 
    numbers=left,                    
    numbersep=5pt,                  
    showspaces=false,                
    showstringspaces=false,
    showtabs=false,                  
    tabsize=2
}

\begin{document}
\title{Archiving and Replaying Current Web Advertisements: Challenges and Opportunities}

\author{Travis Reid} \orcid{0000-0003-1360-7963} \affiliation{ \department{Department of Computer Science} \institution{Old Dominion University} \city{Norfolk} \state{VA} \postcode{23529} \country{USA} } \email{treid003@odu.edu} 

\author{Alex H. Poole} \affiliation{ \department{Department of Information Science} \institution{Drexel University} \city{Philadelphia} \state{PA} \postcode{19104} \country{USA} } \email{ahp56@drexel.edu}

\author{Hyung Wook Choi} \orcid{0000-0002-4075-0768} \affiliation{ \department{Department of Information Science} \institution{Drexel University} \city{Philadelphia} \state{PA} \postcode{19104} \country{USA} } \email{hc685@drexel.edu} 

\author{Christopher Rauch} \affiliation{ \department{Department of Information Science} \institution{Drexel University} \city{Philadelphia} \state{PA} \postcode{19104} \country{USA} } \email{cr625@drexel.edu}  

\author{Mat Kelly} \orcid{0000-0002-0236-7389} \affiliation{ \department{Department of Information Science} \institution{Drexel University} \city{Philadelphia} \state{PA} \postcode{19104} \country{USA} } \email{mkelly@drexel.edu} 

\author{Michael L. Nelson} \orcid{0000-0003-3749-8116} \affiliation{ \department{Department of Computer Science} \institution{Old Dominion University} \city{Norfolk} \state{VA} \postcode{23529} \country{USA} } \email{mln@cs.odu.edu} 

\author{Michele C. Weigle} \orcid{0000-0002-2787-7166} \affiliation{ \department{Department of Computer Science} \institution{Old Dominion University} \city{Norfolk} \state{VA} \postcode{23529} \country{USA} } \email{mweigle@cs.odu.edu}

\begin{abstract}
Although web advertisements represent an inimitable part of digital cultural heritage, serious archiving and replay challenges persist.
To explore these challenges, we created a dataset of 279 archived ads. We encountered five problems in archiving and replaying them.
For one, prior to August 2023, Internet Archive’s Save Page Now service excluded not only well-known ad services' ads, but also URLs with ad related file and directory names. Although after August 2023, Save Page Now still blocked the archiving of ads loaded on a web page, it permitted the archiving of an ad’s resources if the user directly archived the URL(s) associated with the ad. 
Second, Brozzler’s incompatibility with Chrome prevented ads from being archived.
Third, during crawling and replay sessions, Google's and Amazon's ad scripts generated URLs with different random values. This precluded archived ads' replay. Updating replay systems' fuzzy matching approach should enable the replay of these ads.
Fourth, when loading Flashtalking web page ads outside of ad iframes, the ad script requested a non-existent URL. This, prevented the replay of ad resources. But as was the case with Google and Amazon ads, updating replay systems' fuzzy matching approach should enable Flashtalking ads' replay.
Finally, successful replay of ads loaded in iframes with the \texttt{src} attribute of ``\texttt{about:blank}'' depended upon a given browser's service worker implementation. A Chromium bug stopped service workers from accessing resources inside of this type of iframe, which in turn prevented replay. Replacing the ``\texttt{about:blank}'' value for the iframe's \texttt{src} attribute with a blob URL before an ad was loaded solved this problem.
Resolving these replay problems will improve the replay of ads and other dynamically loaded embedded web resources that use random values or ``\texttt{about:blank}'' iframes.
\end{abstract}

\maketitle

\section{Introduction}
Brewster Kahle, founder of the Internet Archive, noted \cite{Kahle-sa97} as early as 1997 that the web constituted ``a storehouse of valuable scientific, cultural and historical information'' (p. 82). Almost a quarter century later, Webster \cite{webster-dpc20} characterized the web in similar terms, but also stressed its potential as ``a vast but underused scholarly resource for the study of almost every possible aspect of the last two decades'' (p. 1). Despite calls to action from numerous scholars, however, web content has been hemorrhaged \cite{cohen-dc10, nelson-arxiv12, pennock-dpc13}. 

Whether impelled by legal obligation, business purposes (e.g., marketing), social or cultural interest, or scholarly and/or historical research, web archiving involves collecting, storing, preserving, and providing long-term access to content \cite{ball-dcc10, cook-ala18, niu-dlib12, pennock-dpc13}. Web archives may be used as evidence about the activities of its creator(s), its users, or about the period in which the archived content was created or modified. 

Because the web depends upon advertising revenue \cite{hwang20}, web ads constitute a particularly important type of dynamic content. Just as physical ephemera in libraries, archives, and museums undergird compelling research, so do online ads illuminate not only the contemporary objectives of advertisers, but also social norms, values, and ideals in ways that curated news stories cannot. Advertisements provide foundational source material for political, social, cultural, and business scholarship, especially in unpacking research questions concerning race, ethnicity, gender, and socioeconomic class \cite{cohen-vb04, ewen-bb01, leach-vb94, marchand-cp86, packard-lgc57}. According to historian Jackson Lears \cite{lears-bb93}, advertisements validate certain worldviews and structures and marginalize others. Advertising, he contends, promotes ``the dominant aspirations, anxieties, even notions of personal identity, in the modern United States'' (p. 2).

Despite web ads' importance, little has been done to build collections of web ads. In large measure, this gap results from the novel and complex technical work it demands.
This exploratory research therefore addresses the following question: what are the key obstacles to archiving and replaying web ads? 
First, we define and situate the core concepts that underpin our work. Next, we explain our methods for archiving, replaying, and identifying resources associated with the 279 ads from our dataset, as well as our tool we developed to find difficult to replay ads. Finally, we describe the five technical challenges we encountered in archiving and replaying web ads. The first problem was the Internet Archive's Save Page Now \cite{ia-savePageNow} service excluding ads from being archived. Second, Brozzler \cite{Levitt-Brozzler-gh14} became incompatible with Chrome after March 2023, which prevented ads from being archived. Third, Google's and Amazon's ad script generated a URL with a random value that differed during the crawling and replay sessions, because replay systems overwrote the random number generator's seed. Fourth, the JavaScript for Flashtalking web page ads requested an unarchived URL when an ad was loaded outside of an iframe, which prevented the resources from loading. Fifth, the ability to replay an ad depended on the browser because the implementation of service workers differed among browsers.

\section{Background}
To understand the archiving and replay problems we identified, we first describe terms and concepts related to web archive crawlers, replay systems, and loading web advertisements on the live web.

\subsection{Web Archive Crawlers and Replay Systems}
\begin{figure}[tbp]
    \centering
     \begin{subfigure}[b]{1.0\textwidth}
        \centering
         \includegraphics[width=\textwidth]{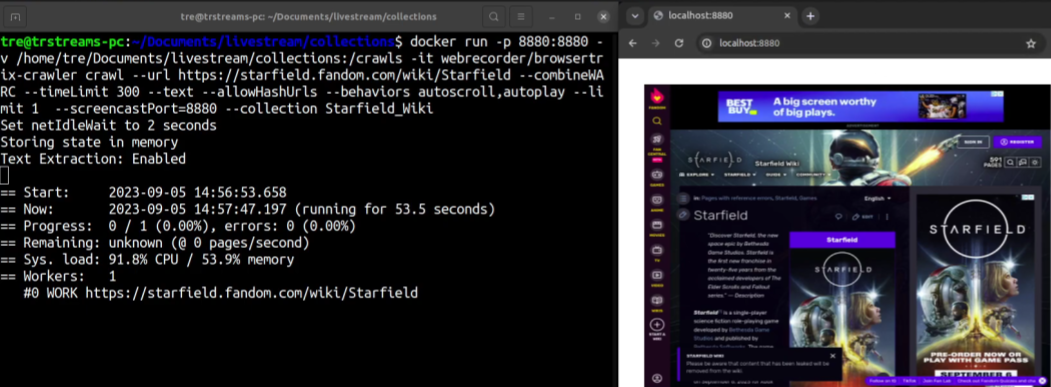}
         \caption{A crawling session where Browsertrix Crawler is archiving a web page}
         \label{fig:crawlingSessionExample}        
     \end{subfigure}
     \vfill
     \begin{subfigure}[b]{1.0\textwidth}
        \centering
         \includegraphics[width=\textwidth]{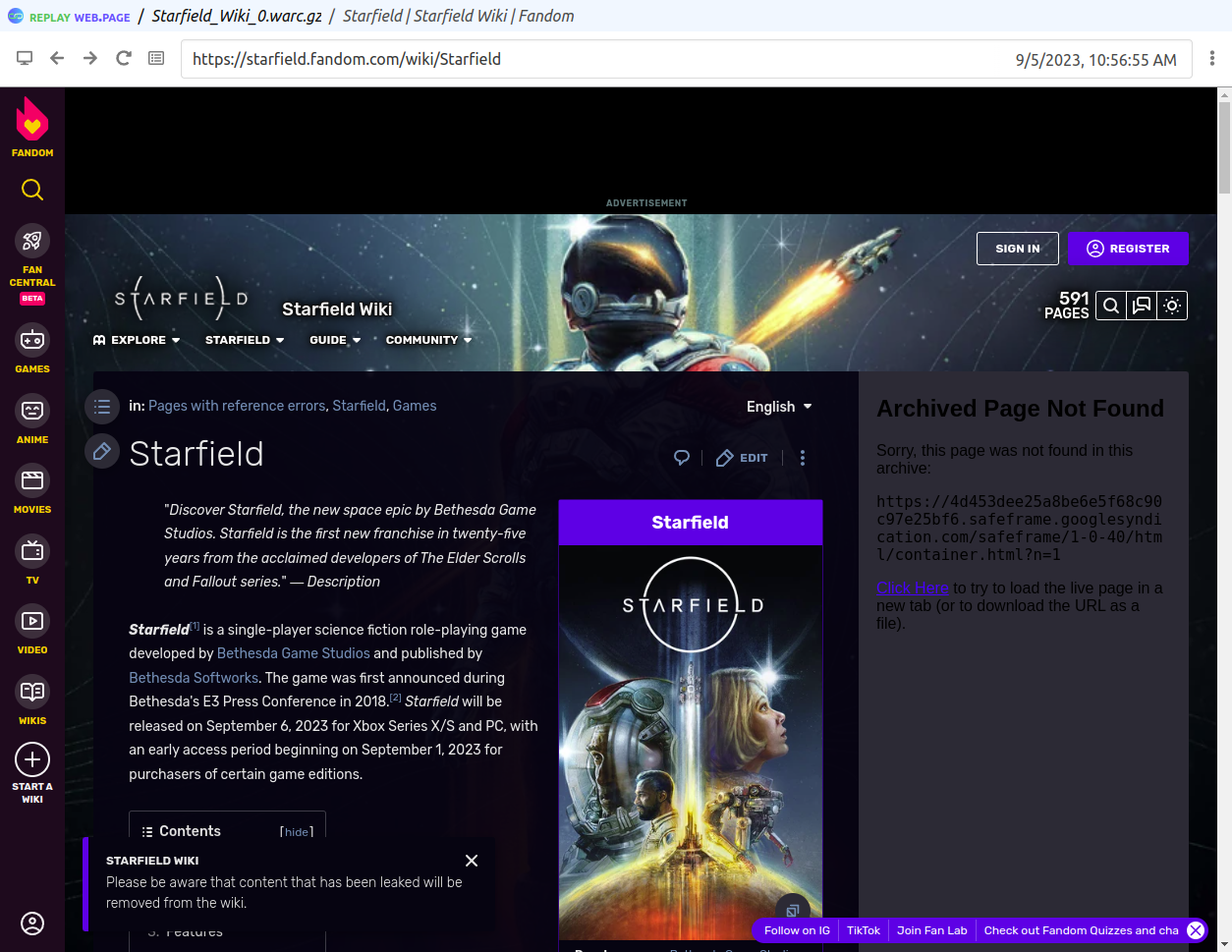}
         \caption{A replay session where ReplayWeb.page is replaying the web page archived by Browsertrix Crawler}
         \label{fig:replaySessionExample}        
     \end{subfigure}
    \caption{Example of a crawling session and replay session.}
    \label{fig:crawlingAndReplaySession}
\end{figure}

\begin{figure}[tbp]
    \centering
     \begin{subfigure}[b]{0.75\textwidth}
        \centering
         \includegraphics[width=\textwidth]{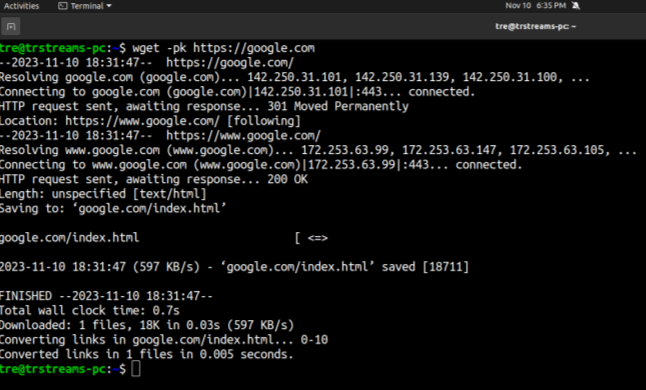}
         \caption{wget is a web crawler that does \emph{not} use a web browser when archiving web pages}
         \label{fig:wget}        
     \end{subfigure}
     \vfill
     \begin{subfigure}[b]{0.75\textwidth}
        \centering
         \includegraphics[width=\textwidth]{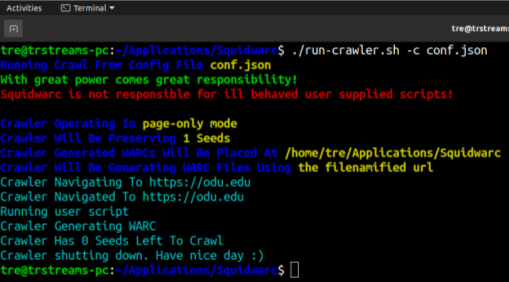}
         \caption{Squidwarc is a web crawler that can use a headless web browser when archiving web pages}
         \label{fig:squidwarc}
     \end{subfigure}
     \vfill
     \begin{subfigure}[b]{0.75\textwidth}
        \centering
         \includegraphics[width=\textwidth]{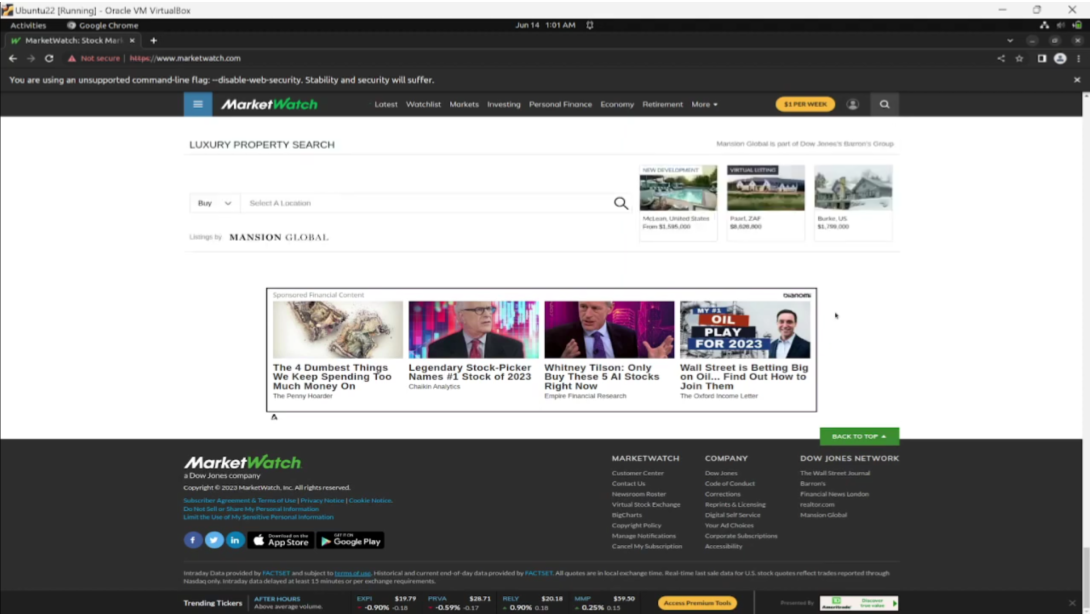}
         \caption{Brozzler is a web crawler that can use a regular web browser (Google Chrome) when archiving web pages}
         \label{fig:Brozzler}
     \end{subfigure}
    \caption{Examples of web crawlers}
    \label{fig:exampleCrawlers}
\end{figure}

Web archiving involves using a crawler to collect content from the World Wide Web and preserve it in an archival format such as WARC (Web ARChive) \cite{warc-iipc13} or WACZ (Web Archive Collection Zipped) \cite{wacz-wr21}. An example crawling session is shown in Figure \ref{fig:crawlingSessionExample}. After the content is archived, a web archive replay system can display the archived version of the web resource(s) in a web browser.
Some web archive crawlers use a graphical user interface (GUI)-based web browser, such as Chrome (bottom of Figure \ref{fig:exampleCrawlers}), or a headless browser, which operates without a GUI (top of Figure \ref{fig:exampleCrawlers}). 

Replaying an archived web page involves using a web archive replay system to load the archived content in a browser, allowing users to view a previously stored version. An example is shown in Figure \ref{fig:replaySessionExample}. 
Replay systems use a URL rewriting system like Wombat \cite{wombat-github23} to modify URLs referenced by an archived webpage's HTML, CSS, and JavaScript files. This rewriting (Listings \ref{urlRewriteLiveWebPage} and \ref{urlRewriteArchivedWebPage}) changes a URI-R (the URI of an Original Resource [i.e., the state of a web resource on the live web at the time it was archived \cite{memento-framework23}]) to a URI-M (the URI for a memento [a previous state of an Original Resource]). In doing so, it prevents live web resources from loading during replay. 
An example URI-M from Wayback Machine is \url{https://web.archive.org/web/20221220095518/https://www.google.com/}. 
This URI-M includes the datetime indicating when the web page was archived (the \textit{Memento-Datetime}): ``\texttt{20221220095518}'' (2022-12-20T09:55:18Z) and the URI-R for the archived web page: ``\url{https://www.google.com/}".
Some web archives, such as Perma.cc~\cite{dulin-permacc-dlib17} and Archive.today~\cite{nelson-wsdlBlog13}, do not include the datetime or the URI-R in their URI-Ms (example URI-M for \url{https://www.google.com/} from Perma.cc is \url{https://perma.cc/V2KT-MYA6} and from Archive.today is \url{https://archive.is/SSsjK}).

Some replay systems, such as ReplayWeb.page, use service workers \cite{archibald-w3c22} to perform client-side URL rewriting. Service workers intercept HTTP requests made by archived web pages during replay \cite{Alam-ServiceWorker-jcdl17, berlin-acm23, berlin-odu18}. In contrast, server-side URL rewriting modifies the URLs in archived HTML, CSS, and JavaScript files before it sends the archived resources to the user. 
This type of rewriting is useful for rewriting URLs in HTML and CSS files. By contrast, in client-side URL rewriting, a service worker or a client-side rewriting JavaScript library like Wombat changes the URL \cite{berlin-odu18}. Wombat, for example, performs the same URL rewriting that is done server-side and overrides the JavaScript API \cite{berlin-odu18}. Client-side URL rewriting outperforms server-side URL rewriting for URLs dynamically generated by JavaScript.

\noindent\begin{minipage}[tb]{\textwidth}
\begin{lstlisting}[language=html, breaklines=false, label=urlRewriteLiveWebPage, caption=Script element before URL rewriting]
<script src="https://treid003.github.io/displayAds.js"> </script>
\end{lstlisting}
\end{minipage}
\noindent\begin{minipage}[tb]{\textwidth}
\begin{lstlisting}[language=html, breaklines=true, label=urlRewriteArchivedWebPage, caption=Script element after URL rewriting]
<script src="https://web.archive.org/web/20240524092904js_/https://treid003.github.io/displayAds.js"> </script>
\end{lstlisting}
\end{minipage}

\subsection{Loading Web Advertisements on the Live Web} \label{Google_SafeFrame_Section}
\begin{figure}[tbp]
    \centering
    \includegraphics[width=\textwidth]{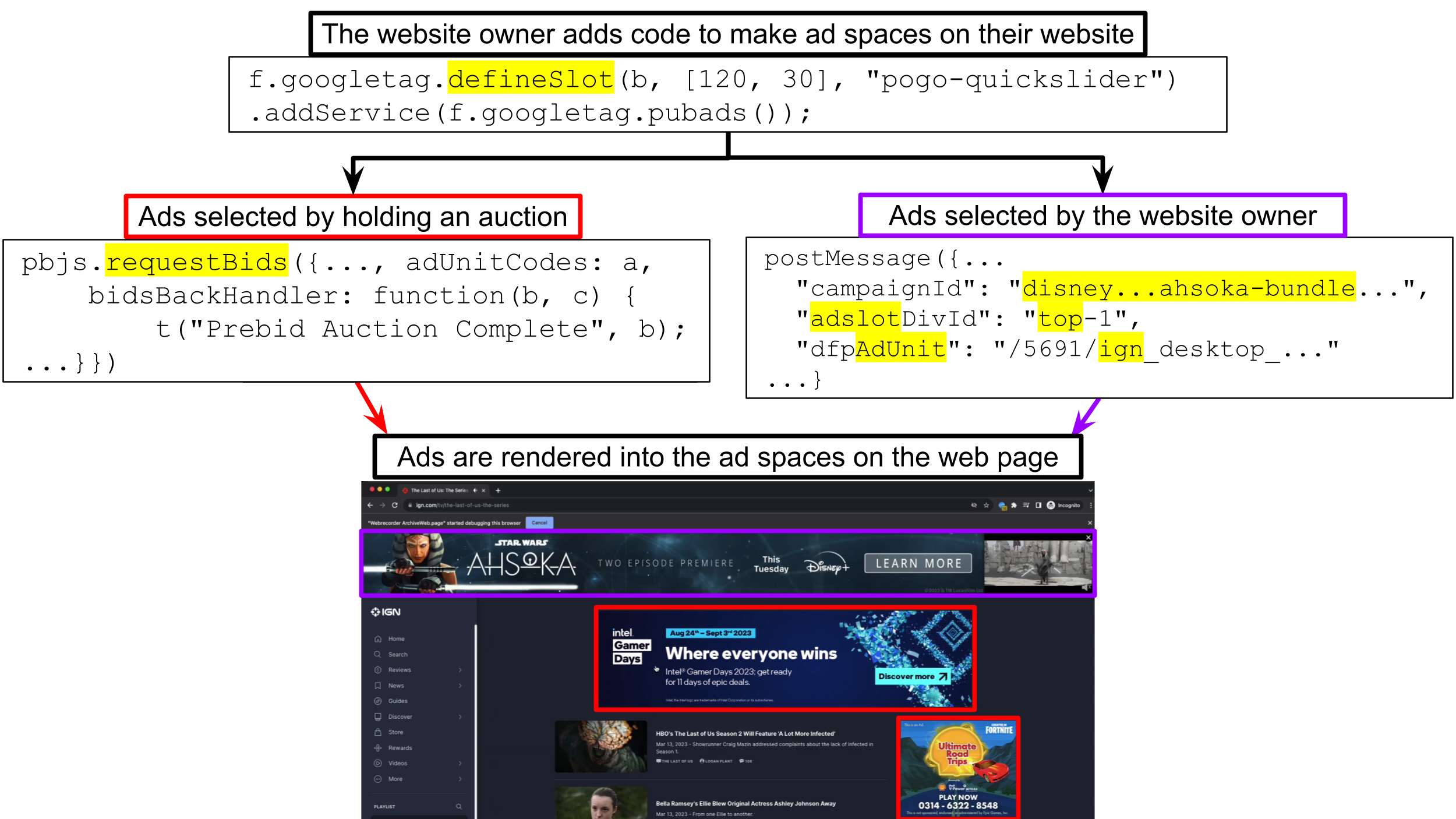}
    \caption{Process for loading advertisements into a web page. IGN's web page (\url{https://www.ign.com/tv/the-last-of-us-the-series}) was used for this example. The code at the top is from \url{https://cdn.ziffstatic.com/pg/ign.js}. the leftmost code is from \url{https://cdn.ziffstatic.com/pg/ign.js}, and the rightmost code is a post message that occurred after IGN requested Disney's ad. WACZ: \url{https://zenodo.org/records/10373131/files/safeframe-example.wacz?download=1}}
    \label{fig:loadWebAds}
\end{figure}

\begin{figure}[tbp]
    \centering
    \includegraphics[width=\textwidth]{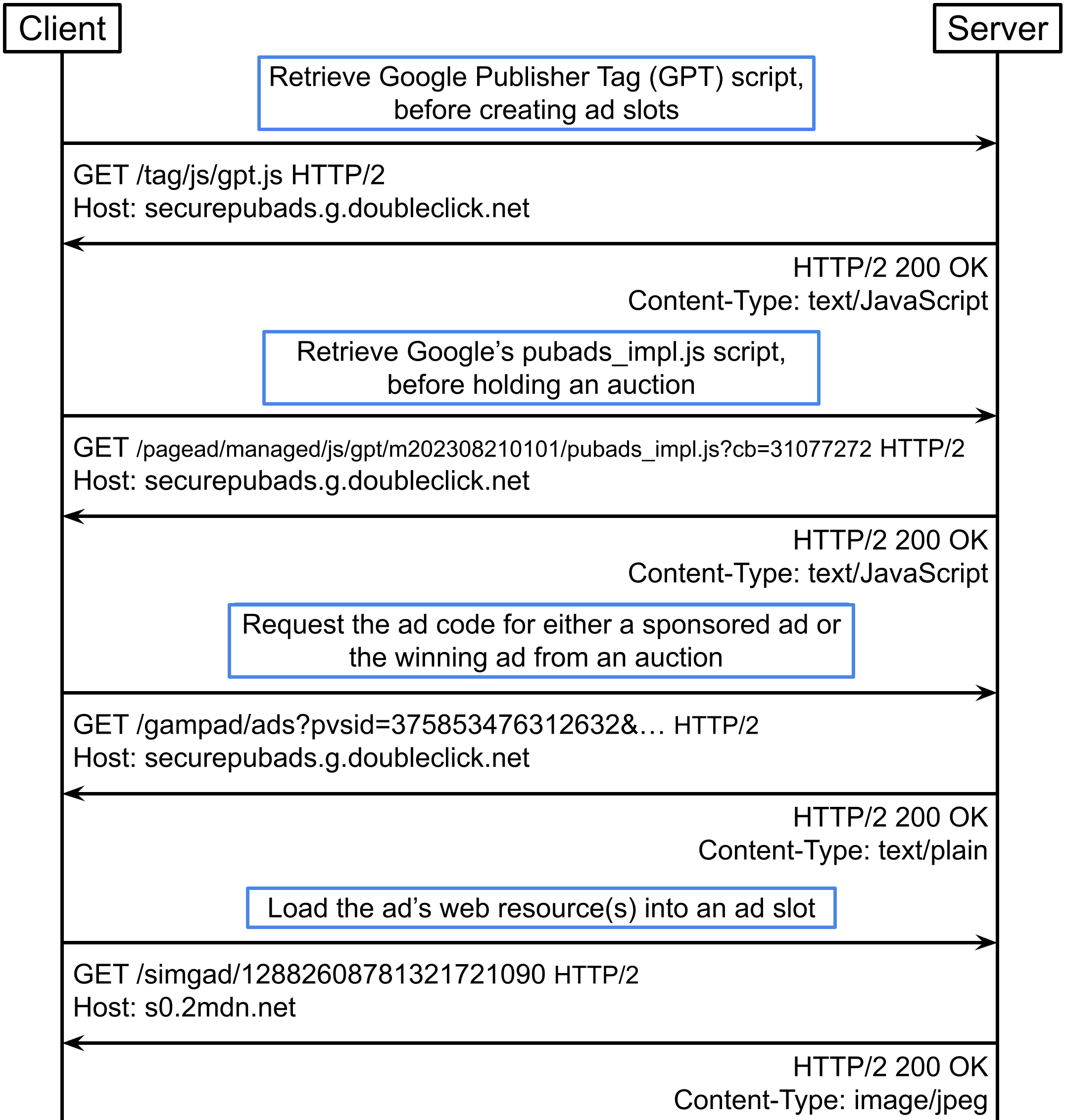}
    \caption{HTTP requests used when loading the Fortnite image ad that was shown in Figure \ref{fig:loadWebAds}.}
    \label{fig:requestsWhenloadingWebAds}
\end{figure}

The process of loading web advertisements involves three steps, as illustrated in Figures \ref{fig:loadWebAds} and \ref{fig:requestsWhenloadingWebAds}. 
First, the publisher (website owner) adds advertisement code, which employs HTML elements such as div and iframe, to create ad spaces~\footnote{Other terms denoting ad space include ad unit \cite{adunit-google24}, ad slot \cite{ad-glossary-google24}, and ad inventory \cite{admanager-google24}.} on their website \cite{adsense-google24}. 
Example code for creating an ad slot is shown at the top row of Figure \ref{fig:loadWebAds}.
During this step, the JavaScript files needed to use the ad service's API, like Google's gpt.js~\footnote{\url{https://securepubads.g.doubleclick.net/tag/js/gpt.js}} and pubads\_impl.js~\footnote{Archived version: \url{https://web.archive.org/web/20230821142108id_/https://securepubads.g.doubleclick.net/pagead/managed/js/gpt/m202308210101/pubads_impl.js?cb=31077272}}, are loaded. The first and second requests shown in Figure \ref{fig:requestsWhenloadingWebAds} were required to use Google Publisher Tag API.
Second, the publisher selects the ad(s) to load either by holding an auction \cite{adsense-google24, bids-amazon24} (example code for requesting auction bids is shown on the left side of the second row of Figure \ref{fig:loadWebAds}) or agreeing to host a sponsored ad \cite{ad-transactions-google24} (the Ashoka-sponsored ad selected by IGN is shown on the right side of the second row of Figure \ref{fig:loadWebAds}).
The ad service selects an auction's winning bid based on metrics such as cost per mile (CPM) (the cost per 1,000 ad impressions~\cite{cpm-amazon24, cpm-google24}).
In the final step, the ad script dynamically retrieves and renders the selected ad(s) into designated ad space(s) on the web page, ensuring proper display and interaction capabilities. The third and fourth HTTP requests in Figure \ref{fig:requestsWhenloadingWebAds} were used to retrieve the code and web resources needed to load the advertisement into an ad slot.
In Figure \ref{fig:loadWebAds}, the Intel and Fortnite ads were selected through auctions, while the website publisher (IGN) selected the Ashoka-sponsored ad.

\begin{figure}[tbp]
    \centering
    \includegraphics[width=\textwidth]{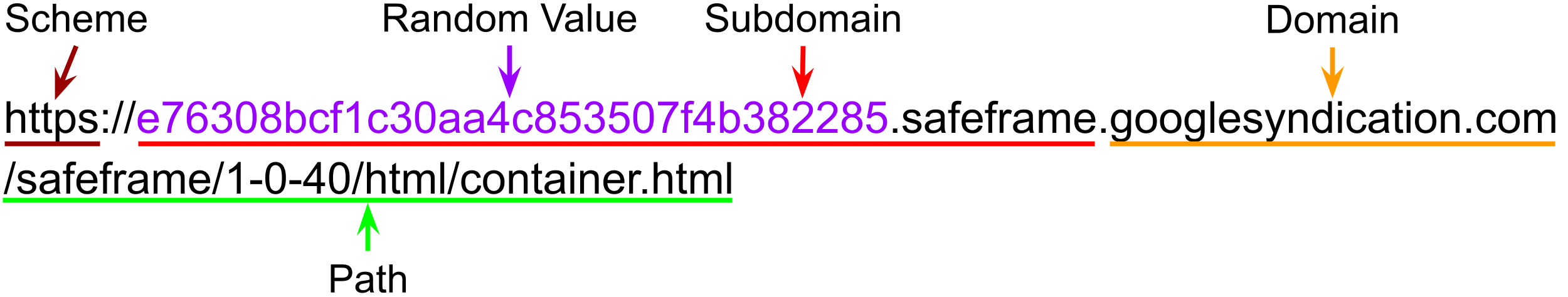}
    \caption{Example URI for a Google SafeFrame. The subdomain contains a random value that is dynamically generated when loading an ad. }
    \label{fig:safeFrameURI}
\end{figure}

\begin{figure}[tbp]
    \centering
    \includegraphics[width=\textwidth]{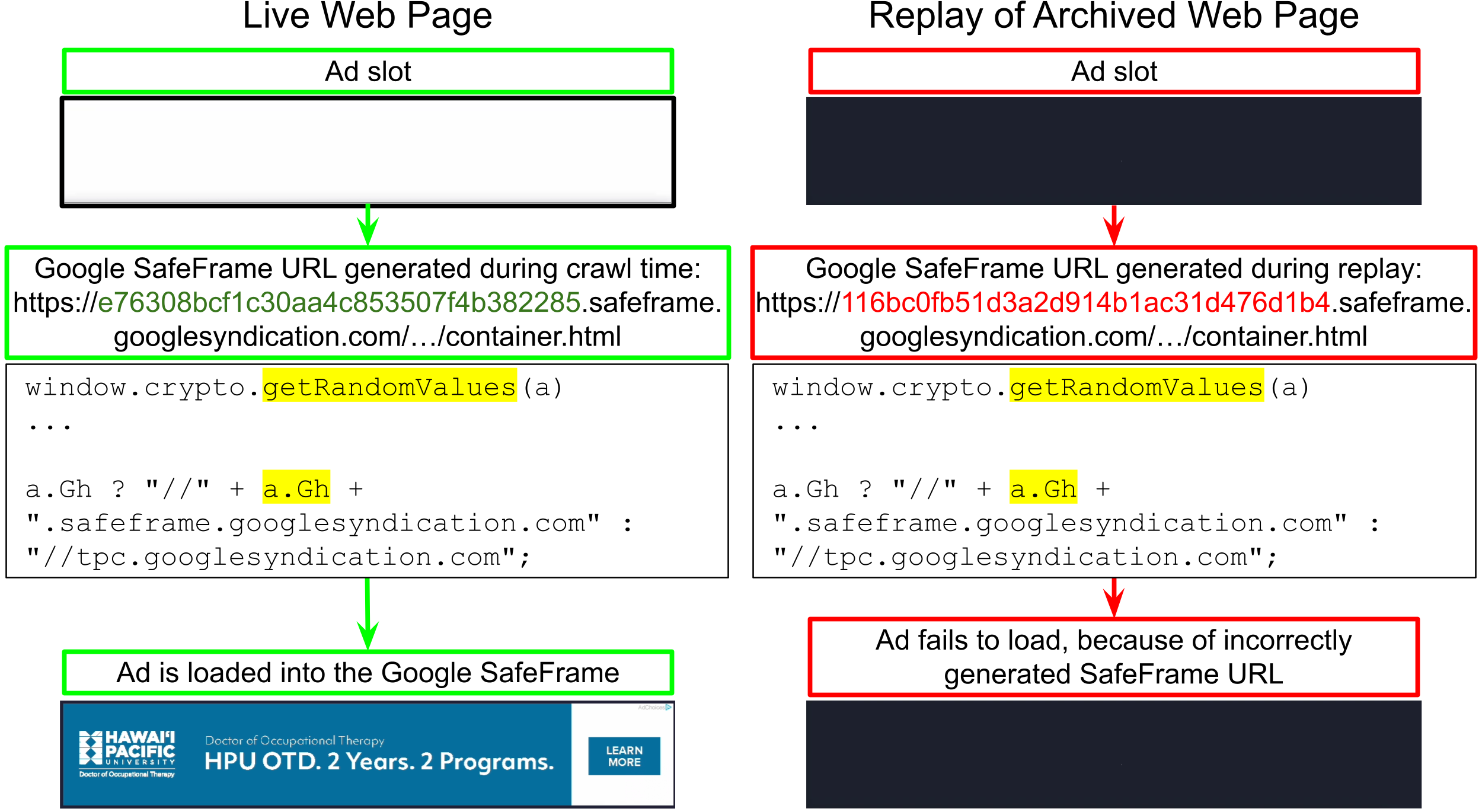}
    \caption{Different SafeFrame URLs during crawl and replay sessions. Google's pubads\_impl.js (URI-R: \url{https://securepubads.g.doubleclick.net/pagead/managed/js/gpt/m202308210101/pubads_impl.js?cb=31077272} | 
    WACZ: \url{https://zenodo.org/records/10373131/files/safeframe-example.wacz?download=1}) generates the random SafeFrame URL.}
    \label{fig:googleAdLiveVsReplay}
\end{figure}

Some Google ads, such as embedded web page ads, use SafeFrame \cite{safeframe-overview-google24}---an iframe based on Interactive Advertising Bureau (IAB) specifications \cite{safeFrame-iab24}---to enhance communication and security between the ad and the hosting web page. Unlike regular iframes, SafeFrames enable controlled communication between the publisher's web page and the ad, while limiting potentially harmful interactions \cite{safeFrame-iab24}.
There are two types of Google SafeFrames. One type does not include a random value in the URL. 
However, this type of Google SafeFrame is disallowed, since Google Publisher Tag's (GPT) API's SafeFrame configuration option \texttt{useUniqueDomain} \cite{uniqeDomain-gpt} is deprecated \cite{gpt-overview-google24, gptAPI-gpt24}.
In contrast, the other type of SafeFrame includes a random value in the URL's subdomain. This sequesters SafeFrame content, thereby strengthening security measures \cite{safeframe-overview-google24}.  Figure \ref{fig:safeFrameURI} shows an example Google SafeFrame URL with a dynamically generated random subdomain. 
This type of SafeFrame remains difficult to replay because the random value generated during replay by Google's \href{https://web.archive.org/web/20230113005605id_/https://securepubads.g.doubleclick.net/gpt/pubads_impl_2023010901.js?cb=31071543}{pubads\_impl\_2023020201.js}~\footnote{URI-M: \url{https://web.archive.org/web/20230113005605id_/https://securepubads.g.doubleclick.net/gpt/pubads_impl_2023010901.js?cb=31071543}} script will differ from the random value generated at crawl time. This replay problem will be discussed in more detail in Section \ref{Replaying_Google_SafeFrame}.
Figure \ref{fig:googleAdLiveVsReplay} shows an example ad that failed to load because of this problem. 

\begin{figure}[tbp]
    \centering
    \includegraphics[width=\textwidth]{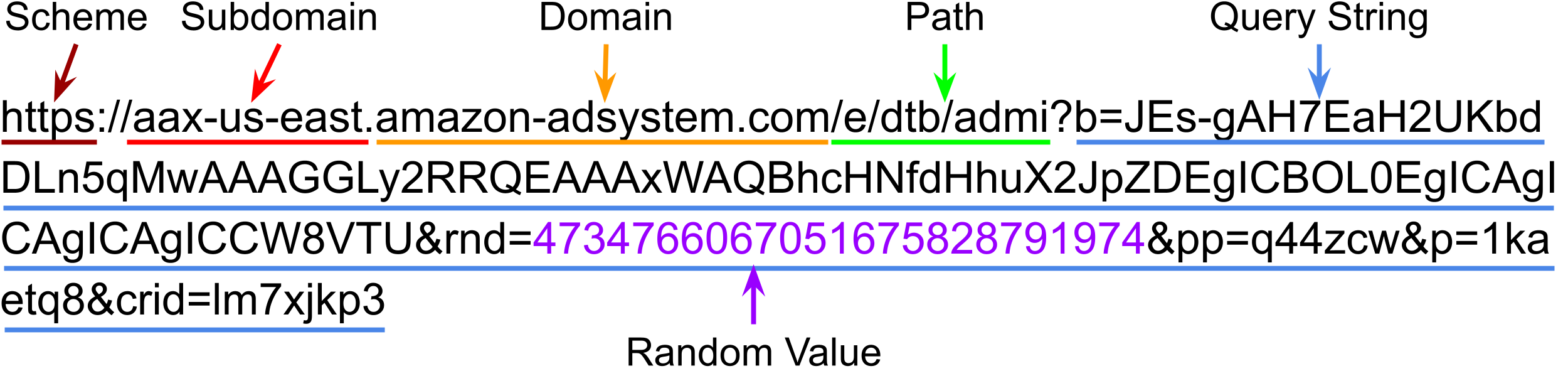}
    \caption{Example URI for an Amazon ad iframe. The \texttt{rnd} parameter in the query string contains a random value that is dynamically generated when loading an ad.}
    \label{fig:amazonAdIframeURI}
\end{figure}

Like Google's SafeFrame, Amazon's ad iframe uses a random value in the iframe's URL, albeit one  located in the query string (Figure \ref{fig:amazonAdIframeURI}) instead of the subdomain. 
A random value's presence in the query string results in the replay system generating an unarchived Amazon ad iframe URL. This can prevent an Amazon ad from loading during replay.

\section{Methods}
In this work, we created a dataset~\footnote{We created a web page to display ads from our dataset: \url{https://savingads.github.io/themed_ad_collections.html}} \cite{ad-dataset-gh24} of 279 unique recent (January-June, 2023) advertisements culled from the live web. We archived them with appropriate tools, and identified problems with both their archiving and  their replay.

\subsection{Creating a Dataset of Recently Archived Ads} \label{section:creatingDataset}
To construct our dataset, we randomly selected websites from Similarweb's top websites worldwide \cite{similarweb} (including all categories except ``Adult''~\footnote{The ``Adult'' category was excluded because we planned on recording and uploading videos of the crawling and replay sessions to YouTube. Websites from this category cannot be included in a video, as it would violate YouTube's Community Guidelines (\url{https://support.google.com/youtube/answer/2802002?hl=en&ref_topic=9282679}).}), rendered a web page from each website, and if the page loaded ads, archived it. We repeated this process until we had collected at least 250 ads.

\begin{table}[tbp]
\begin{tabular}{|l|r|l|l|}
\hline
\multicolumn{1}{|c|}{\textbf{Web Page}}                                                           & \multicolumn{1}{c|}{\textbf{\begin{tabular}[c]{@{}c@{}}Number of \\ Ads Archived\end{tabular}}} & \multicolumn{1}{c|}{\textbf{\begin{tabular}[c]{@{}c@{}}Web Archiving \\ Tool\end{tabular}}} & \multicolumn{1}{c|}{\textbf{Replay System}} \\ \hline
\href{https://canalturf.com}{https://canalturf.com}                                                                             & 66                                                                                             & Save Page Now                                                                               & Wayback Machine                             \\ \hline
\begin{tabular}[c]{@{}l@{}}\href{https://www.lequipe.fr/Tous-sports/Actualites/Le-flash-sports-du-5-avril/1389820}{https://www.lequipe.fr}\\ \href{https://www.lequipe.fr/Tous-sports/Actualites/Le-flash-sports-du-5-avril/1389820}{/Tous-sports/.../1389820}\end{tabular}         & 37                                                                                              & ArchiveWeb.page                                                                             & ReplayWeb.page                              \\ \hline
\begin{tabular}[c]{@{}l@{}}\href{https://www.leroymerlin.com.br/}{https://www.leroymerlin}\\ \href{https://www.leroymerlin.com.br/}{.com.br}/\end{tabular}                        & 31                                                                                              & Arquivo.pt                                                                                  & Arquivo.pt                                  \\ \hline
\begin{tabular}[c]{@{}l@{}}\href{https://www.ign.com/articles/the-last-of-us-season-1-review}{https://www.ign.com}\\ \href{https://www.ign.com/articles/the-last-of-us-season-1-review}{/articles/the-last-...-review}\end{tabular}       & 24                                                                                              & ArchiveWeb.page                                                                             & ReplayWeb.page                              \\ \hline
\begin{tabular}[c]{@{}l@{}}\href{https://www.facebook.com/}{https://www.facebook}\\ \href{https://www.facebook.com/}{.com/}\end{tabular}                              & 24                                                                                              & ArchiveWeb.page                                                                             & ReplayWeb.page                              \\ \hline
\url{https://www.cnn.com/}                                                                              & 23                                                                                              & ArchiveWeb.page                                                                             & ReplayWeb.page                              \\ \hline
\begin{tabular}[c]{@{}l@{}}\href{https://www.marketwatch.com/}{https://www.marketwatch}\\ \href{https://www.marketwatch.com/}{.com/}\end{tabular}                           & 18                                                                                              & Brozzler                                                                                    & ReplayWeb.page                              \\ \hline
\url{https://www.diy.com/}                                                                              & 13                                                                                              & Conifer                                                                                     & Conifer                                     \\ \hline
\begin{tabular}[c]{@{}l@{}}\href{https://www.realtor.com/news/unique-homes/frank-lloyd-wright-designed-home-in-tulsa-ok-lists-for-7-9m/}{https://www.realtor.com}\\ \href{https://www.realtor.com/news/unique-homes/frank-lloyd-wright-designed-home-in-tulsa-ok-lists-for-7-9m/}{/news/.../frank-...7-9m/}\end{tabular}        & 11                                                                                              & Browsertrix Crawler                                                                         & ReplayWeb.page                              \\ \hline
\begin{tabular}[c]{@{}l@{}}\href{https://mortalkombat.fandom.com/wiki/Tag_Team_Ladder}{https://mortalkombat.fandom}\\ \href{https://mortalkombat.fandom.com/wiki/Tag_Team_Ladder}{.com/wiki/Tag\_Team\_Ladder}\end{tabular} & 8                                                                                               & Browsertrix Crawler                                                                         & ReplayWeb.page                              \\ \hline
\url{https://tokopedia.com}                                                                             & 6                                                                                               & Arquivo.pt                                                                                  & Arquivo.pt                                  \\ \hline
\begin{tabular}[c]{@{}l@{}}\href{https://www.deviantart.com/kvacm/art/Hellstone-Ruins-860415274}{https://www.deviantart.com}\\ \href{https://www.deviantart.com/kvacm/art/Hellstone-Ruins-860415274}{/kvacm/art/Hellstone-...274}\end{tabular}  & 5                                                                                               & Browsertrix Crawler                                                                         & ReplayWeb.page                              \\ \hline
\begin{tabular}[c]{@{}l@{}}\href{https://unsplash.com/t/wallpapers}{https://unsplash.com/t}\\ \href{https://unsplash.com/t/wallpapers}{/wallpapers}\end{tabular}                      & 3                                                                                               & archive.today                                                                               & archive.today                               \\ \hline
\url{https://sports.yahoo.com}                                                                          & 3                                                                                               & Conifer                                                                                     & Conifer                                     \\ \hline
\begin{tabular}[c]{@{}l@{}}\href{https://www.vidal.ru/novosti/kak-potreblenie-razlichnyh-doz-alkogolya-vliyaet-na-smertnost-11744}{https://www.vidal.ru/novosti}\\ \href{https://www.vidal.ru/novosti/kak-potreblenie-razlichnyh-doz-alkogolya-vliyaet-na-smertnost-11744}{/kak-potreblenie-...744}\end{tabular}    & 2                                                                                               & Save Page Now                                                                               & Wayback Machine                             \\ \hline
\url{https://canalturf.com}                                                                             & 2                                                                                               & ArchiveBot                                                                                  & Wayback Machine                             \\ \hline
\url{https://canalturf.com}                                                                             & 1                                                                                               & Perma.cc                                                                                    & Wayback Machine                             \\ \hline
\begin{tabular}[c]{@{}l@{}}\href{https://www.youtube.com/watch?v=PZShwWiepeY}{https://www.youtube.com}\\ \href{https://www.youtube.com/watch?v=PZShwWiepeY}{/watch?v=PZShwWiepeY}\end{tabular}            & 1                                                                                               & Browsertrix Crawler                                                                         & ReplayWeb.page                              \\ \hline
\begin{tabular}[c]{@{}l@{}}\href{https://www.tripadvisor.it/Tourism-g186338-London_England-Vacations.html}{https://www.tripadvisor.it}\\ \href{https://www.tripadvisor.it/Tourism-g186338-London_England-Vacations.html}{/Tourism-...-Vacations.html}\end{tabular}  & 1                                                                                               & archive.today                                                                               & archive.today                               \\ \hline
\end{tabular}
\caption{The number of archived ads from each web page that we archived when creating the dataset.}
\label{table:adsPerWebpage}
\end{table}

Ultimately, we selected 17 web pages and 279 ads to archive (Table \ref{table:adsPerWebpage}). We used Internet Archive's Save Page Now, Arquivo.pt, archive.today, and Conifer because these web archiving services permit the archiving of an unlimited number of web pages cost-free. We also used three browser-based tools (ArchiveWeb.page \cite{kreymer-ArchiveWebpage-gh20}, Browsertrix Crawler, and Brozzler) that facilitate archiving dynamically loaded web resources. ArchiveWeb.page and Browsertrix Crawler archived four web pages each, Brozzler archived one web page, and four web archiving services (Save Page Now~\footnote{When replaying a web page, \url{https://canalturf.com}, archived by Save Page Now, we found that some of the ads that were loaded by Wayback Machine were archived by ArchiveBot and Perma.cc}, Arquivo.pt, archive.today, and Conifer) archived two web pages each. We did not archive four web pages with each web archiving tool because we had reached our goal of 250 advertisements.
We successfully archived (captured all the resources needed) nearly all (273) of these ads.

Six ads were not fully archived. One required a specific user interaction (clicking on a play button) to load all of the ad resources. Three ads requested unarchived JavaScript and HTML files during replay. We used ReplayWeb.page's URL prefix search and removed the query string from the requested URI to see if a resource with a similar URI was archived, but did not find these JavaScript and HTML files in ArchiveWeb.page's output file (WACZ). For one Flashtalking ad, it was impossible to determine if the ad  was successfully archived because this type of ad cannot be replayed outside of its ad iframe. This prevented us from comparing the ad resources that loaded on the live web and (presumably) would load during replay. (This problem with Flashtalking ads will be discussed in Section \ref{section:flashtalking}.) The last partially archived ad's dynamically generated URL (during crawl time) included an  \texttt{e} query string parameter that prevented\footnote{Ad URI with \texttt{e} parameter: \url{https://s0.2mdn.net/sadbundle/14073241274752696320/970x250-HBO_Max/index.html?e=69&leftOffset=0&topOffset=0&c=qipO0hxSN3&t=1&renderingType=2&ev=01_247}}\footnote{Ad URI without \texttt{e} parameter: \url{https://s0.2mdn.net/sadbundle/14073241274752696320/970x250-HBO_Max/index.html?leftOffset=0&topOffset=0&c=qipO0hxSN3&t=1&renderingType=2&ev=01_247}} some of the images from loading. There were seven more images on the live version of the ad, but the live web ad was checked months after crawl time.

\subsubsection{Finding Archived Ads' Web Resources}
By using ReplayWeb.page's URL search feature \cite{replayWebPage-search-24} and our own bespoke Display Archived Ads tool \cite{display-archived-ads-gh24}, we found that 55 out of 279 advertisements were not replayable when loading the archived containing web page~\footnote{A containing web page is the web page that loaded the ad during a crawling session.}. 
We used ReplayWeb.page's URL search feature to identify ads in our WARC and WACZ files. This feature allows users to specify the MIME type, which facilitates searching for a specific ad type like image ads. It also allows for a prefix search, which we used to identify resources associated with services like Flashtalking, Innovid, Amazon display ads, and Google AdSense. 

The steps we used for ReplayWeb.page's URL search are listed below~\footnote{Video example: \url{https://youtu.be/QXUDINMJ5Ec?t=238s}}: \begin{enumerate} \item Load WARC or WACZ file with ReplayWeb.page \item Select ``URLs'' tab \item For the button beside ``Search'', select the MIME type for the ad \item Enter a URL prefix associated with an ad service, such as \url{https://s0.2mdn.net/} \item Click on the search result items to load the resource (the image, video, or HTML file) and see if the resource is an ad \end{enumerate}

\begin{figure}[tbp]
    \centering
    \includegraphics[width=\textwidth]{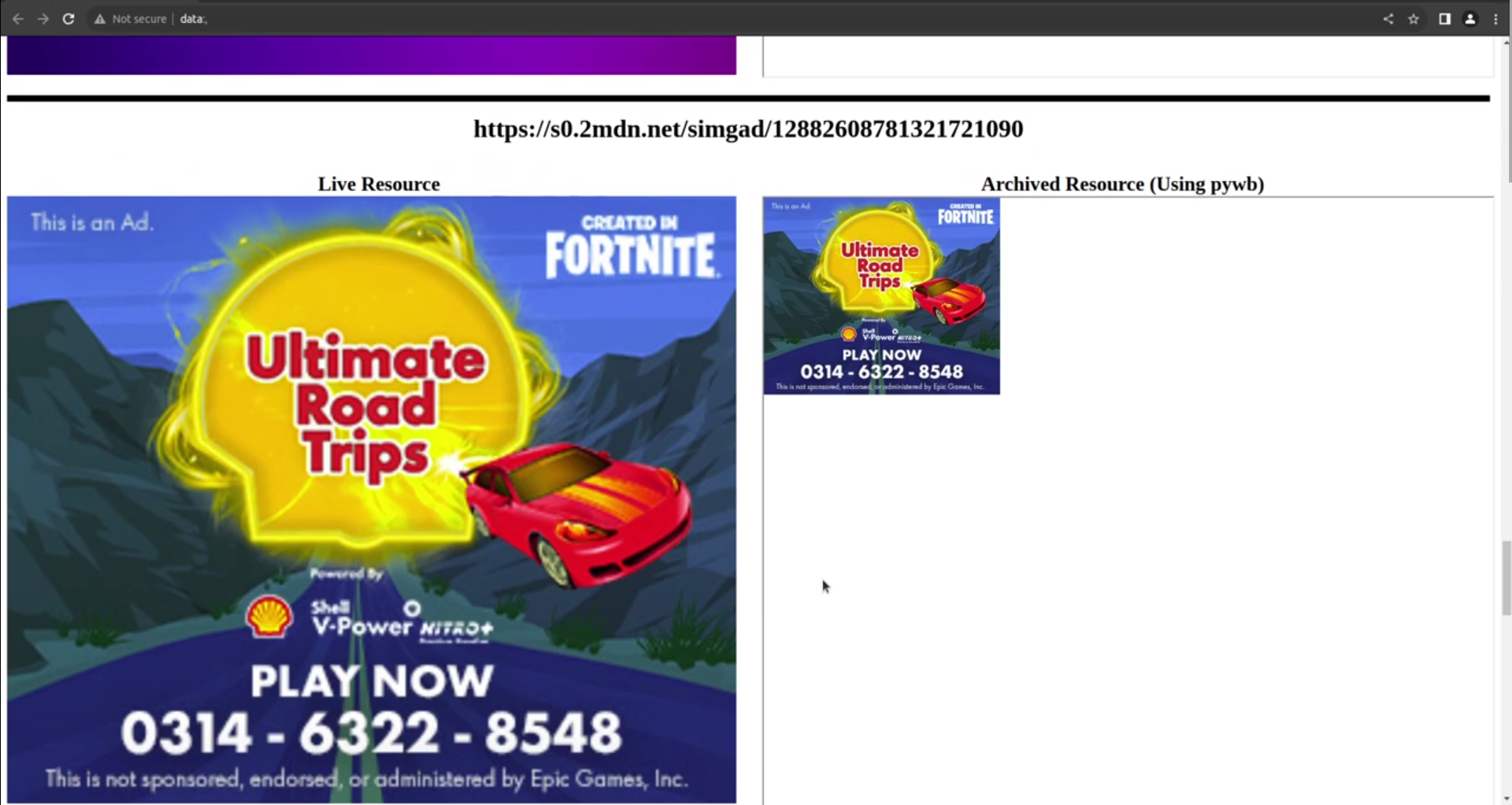}
    \caption{Our tool for displaying potential ads loading the live web ad beside the archived version of the same ad.}
    \label{fig:displayPotentialAds}
\end{figure}

Our Display Archived Ads tool~\footnote{Demo video for our tool: \url{https://youtu.be/Bc2T6ZZd210?t=30}} (Figure \ref{fig:displayPotentialAds}) was used to display most of the HTML, image, and video files inside of a given WARC file.
Our tool depends on the warcio~\cite{kreymer-warcio-gh24}, pywb~\cite{pywb-gh13}, and Selenium~\cite{selenium} software packages. Warcio retrieves the URLs for the web resources from the WARC file. Pywb replays the archived ads within an iframe~\footnote{We replayed the archived resources in an iframe in order to display multiple ads on the same web page.}, and Selenium opens a web browser and executes the JavaScript necessary to display the ads. 
Our tool offered two affordances. 
It enabled us to filter out some of the known ad resources that remain invisible during replay~\footnote{An example filtered file is pixel.gif, a commonly used image for ad services that only shows a few white pixels.}, thereby speeding up the review. Further, by allowing us to display an ad's live version beside its archived version, the tool showed problems with replay.
The command for running our tool requires a WARC file and the seed URL, which is used to exclude the containing web page from the HTML category. Listing \ref{displayAdsGeneral}  shows the general structure for the command. Listing \ref{displayAdsExample} shows an example command for an IGN web page that includes ads.

\noindent\begin{minipage}[tb]{\textwidth} \begin{lstlisting}[language=bash, breaklines=false, label=displayAdsGeneral, caption=General structure for the command used for our tool] 
python3 Display_Archived_Ads.py /path/to/WARC/file.warc <seed URL> \end{lstlisting} \end{minipage}
\noindent\begin{minipage}[tb]{\textwidth} \begin{lstlisting}[language=bash, breaklines=true, label=displayAdsExample, caption=Example command used for an IGN web page] 
python3 Display_Archived_Ads.py data.warc.gz https://www.ign.com/tv/the-last-of-us-the-series \end{lstlisting} \end{minipage}

\subsubsection{Replaying Archived Ads}
To replay the archived advertisements, we used four web archiving services (Internet Archive's Wayback Machine, Arquivo.pt, archive.today, and Conifer). We replayed the web ads that we archived with a web archiving service with a service from the same web archive. We also used three other replay systems (ReplayWeb.page, pywb, and OpenWayback \cite{OpenWayback-gh12}). 
We used ReplayWeb.page to replay the archived ads from our WARC and WACZ files, because at the beginning of 2023, ReplayWeb.page could replay more dynamically loaded web resources than pywb and OpenWayback. Pywb (version 2.7.3) failed to replay archived web ads that relied upon an Amazon ad iframe. During replay, OpenWayback (version 2.4.0) \cite{ruest-owb-240-gh19} loaded live web advertisements instead of the archived ads.  Brunelle \cite{brunelle-zombies-wsdl12} and Lerner et al. \cite{Lerner-acm17} discuss this problem of live web resources being loaded during replay.  

\subsection{Categorizing Advertisements}
\begin{table}[tbp]
\begin{tabular}{|l|r|}
\hline
\multicolumn{1}{|c|}{\textbf{Theme}} & \multicolumn{1}{l|}{\textbf{Number of Ads}} \\ \hline
Shopping                             & 85                                          \\ \hline
Finance                              & 27                                          \\ \hline
Vehicle and Automotive               & 23                                          \\ \hline
Business Services                    & 21                                          \\ \hline
Travel                               & 19                                          \\ \hline
Entertainment                        & 16                                          \\ \hline
Health                               & 15                                          \\ \hline
Real Estate                          & 15                                          \\ \hline
News                                 & 14                                          \\ \hline
Unknown                              & 6                                           \\ \hline
Internet and Mobile Service Provider & 5                                           \\ \hline
Art                                  & 4                                           \\ \hline
Beauty and Cosmetics                 & 4                                           \\ \hline
Gaming                               & 4                                           \\ \hline
Military                             & 4                                           \\ \hline
Food and Drink                       & 3                                           \\ \hline
Gambling and Fantasy Sports          & 3                                           \\ \hline
Computer Security                    & 2                                           \\ \hline
Fitness and Sports                   & 2                                           \\ \hline
Pets and Animals                     & 2                                           \\ \hline
Sponsored Brand                      & 2                                           \\ \hline
Charity                              & 1                                           \\ \hline
Funeral Services                     & 1                                           \\ \hline
Politics                             & 1                                           \\ \hline
\end{tabular}
\caption{List of themes used for the ads in our dataset.}
\label{table:adThemes}
\end{table}

\begin{figure}[tbp]
    \centering
    \includegraphics[width=\textwidth]{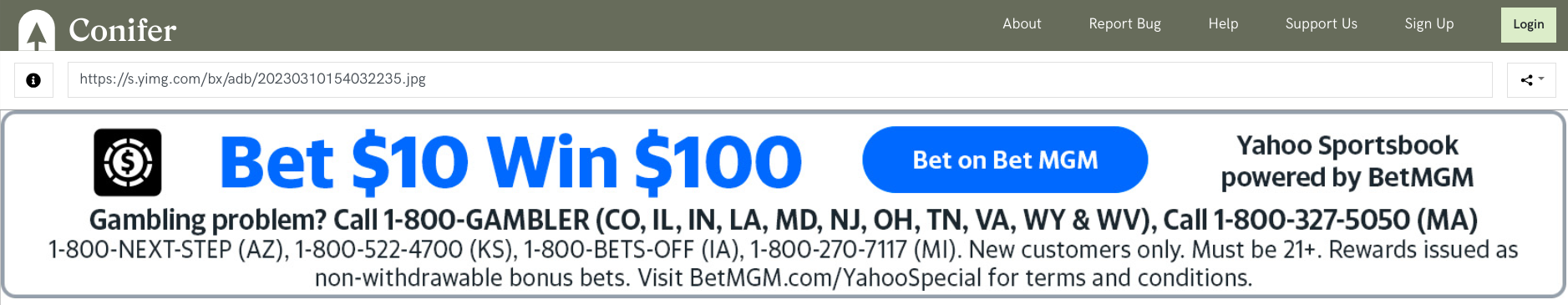}
    \caption{An example image ad loaded outside of the containing web page. Ad's URI-M: \url{https://conifer.rhizome.org/treid003/2023-05-16-archiving-ads-on-sportsyahoocom/https://s.yimg.com/bx/adb/20230310154032235.jpg} }
    \label{fig:imageAd}
\end{figure}

\begin{figure}[tbp]
    \centering
    \includegraphics[scale=0.25]{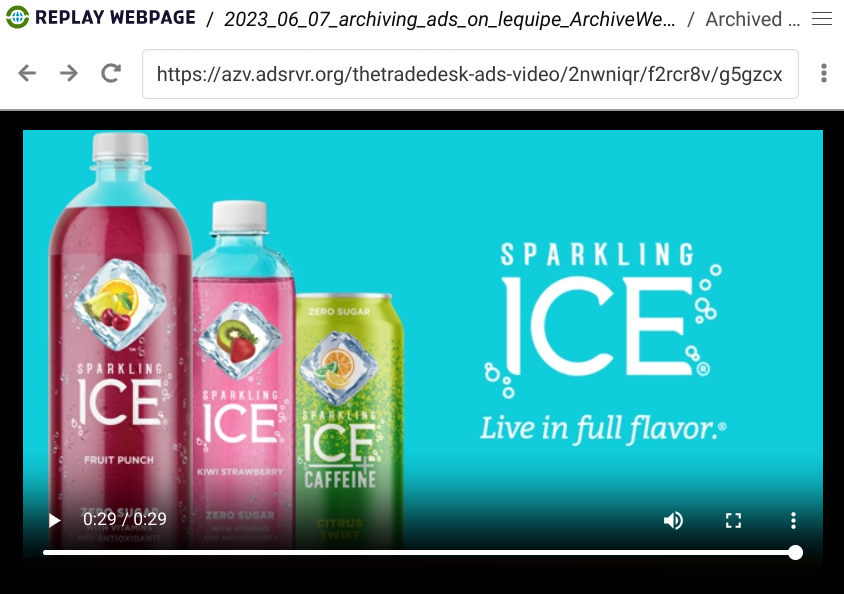}
    \caption{An example video ad loaded outside of the containing web page. WACZ: \url{https://zenodo.org/record/8057942/files/2023_06_07_archiving_ads_on_lequipe_ArchiveWeb_page.wacz?download=1} | Ad's URI-R: \url{https://azv.adsrvr.org/thetradedesk-ads-video/2nwniqr/f2rcr8v/g5gzcxa29115fa8fdeed4ef88089cec513d745e4.mp4} }
    \label{fig:videoAd}
\end{figure}

\begin{figure}[tbp]
    \centering
    \includegraphics[scale=0.25]{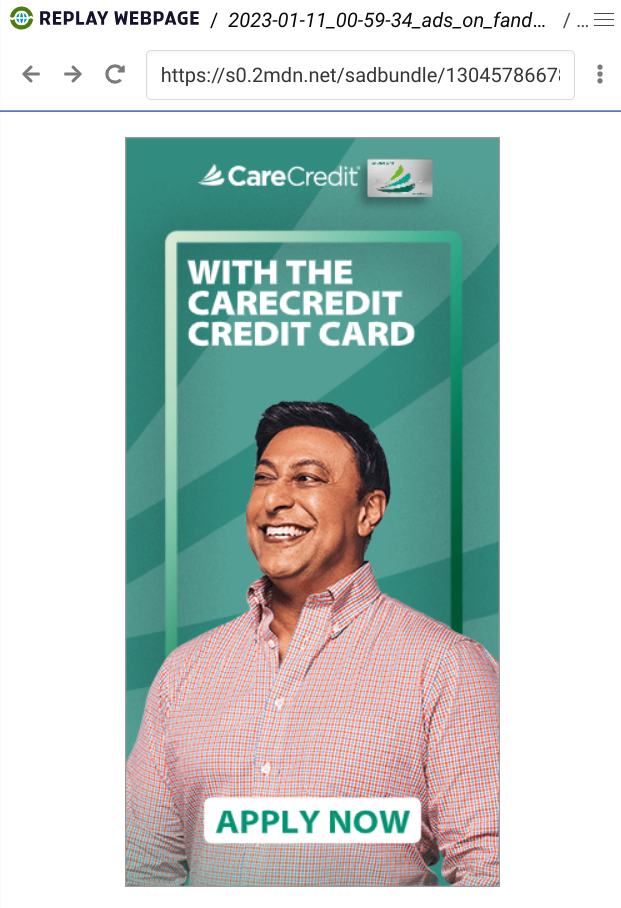}
    \caption{An example web page ad loaded outside of the containing web page. WARC: \url{https://zenodo.org/record/7601187/files/2023-01-11_00-59-34_ads_on_fandom_browsertrix_crawler.warc.gz?download=1} | Ad's URI-R: \url{https://s0.2mdn.net/sadbundle/13045786678919115269/CCD2C_5568424_300x600_MF_CP_APPLY_NA_NR_EN_V1_H5_BD_2022_042025/index.html} }
    \label{fig:embeddedWebPageAd}
\end{figure}

\begin{figure}[tbp]
    \centering
    \includegraphics[scale=0.3]{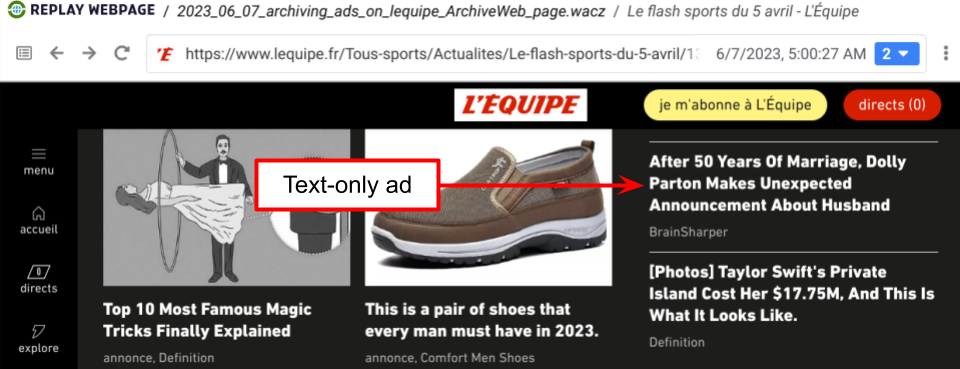}
    \caption{An example text-only ad for a sponsored news article loaded in the containing web page. WACZ: \url{https://zenodo.org/record/8057942/files/2023_06_07_archiving_ads_on_lequipe_ArchiveWeb_page.wacz?download=1} | Containing web page's URI-R: \url{https://www.lequipe.fr/Tous-sports/Actualites/Le-flash-sports-du-5-avril/1389820} }
    \label{fig:textAd}
\end{figure}

\begin{figure}[tbp]
    \centering
    \includegraphics[scale=0.3]{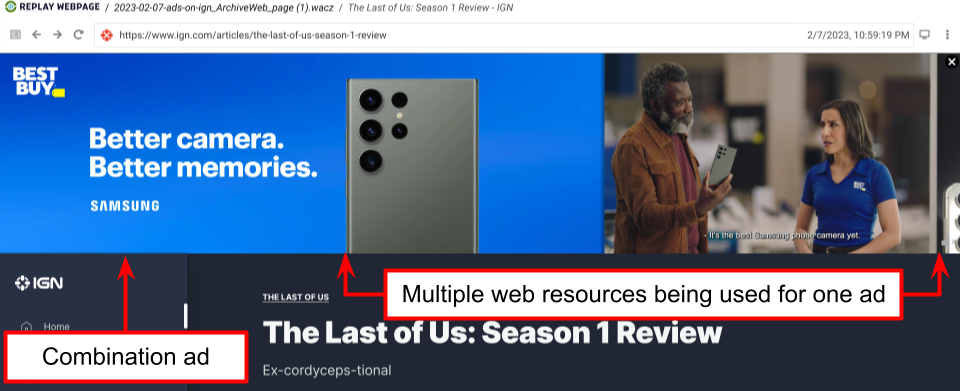}
    \caption{An example combination ad loaded in the containing web page. This ad uses three images and one video. WACZ: \url{https://zenodo.org/record/8000975/files/2023-02-07-ads-on-ign_ArchiveWeb_page.wacz?download=1} | Containing web page's URI-R: \url{https://www.ign.com/articles/the-last-of-us-season-1-review} }
    \label{fig:combinationAd}
\end{figure}

We categorized our 279 ads into five categories: image, video, embedded web page, text-only, or a combination.
The first three types are associated with one web resource that is viewable outside of the containing web page (provided the user knows the URI associated with the resource). Figures \ref{fig:imageAd} (image ad), \ref{fig:videoAd} (video ad), and \ref{fig:embeddedWebPageAd} (embedded web page ad) show examples.
Text-only ads (Figure \ref{fig:textAd}) cannot be viewed outside of the containing web page because the web page loads the text.
The combination category comprises ads (Figure \ref{fig:combinationAd}) that rely upon multiple resources and are constructed inside of the containing web page or ad iframe. Like text ads, combination ads cannot be viewed outside of the containing web page.

Next, we coded each ad topically.
Table \ref{table:adThemes} shows the 24 themes and the corresponding number of ads for each.
Most themes (17 of 24) aligned with SimilarWeb's website categories. The other themes included ``Internet and Mobile Service Provider'', ``Politics'', ``Funeral Services'', ``Charity'', ``Military'', ``Sponsored Brand'', and ``Unknown''\footnote{``Unknown'' refers to ads that we were not able to replay and could not view on the live web.}.
Notably, the Military theme was exclusively video ads. In contrast, the other themes with more than three ads included multiple ad types.
After this thematic coding, we sought to identify the problems with archiving and replaying the 279 ads.

\section{Findings}
We identified five key archiving and replay problems. First, we discuss two archiving problems that involved the Internet Archive's Save Page Now excluding ads and recent versions of Chrome being incompatible with Brozzler. Second, we describe three replay problems caused by random values in URLs, by a non-existent URL being requested, and by a Chromium bug that prevented service workers from accessing resources in an  ``\texttt{about:blank}'' iframe.

\subsection{Archiving Problems}
Two problems prevented web ads from being archived: Internet Archive's Save Page Now intentionally excluding ads and Brozzler being incompatible with recent versions of Google Chrome.

\subsubsection{Internet Archive’s Save Page Now Blocks Ads From Being Archived}
We inspected ads available on Wayback Machine using three URL searches with the prefixes ``https://s0.2mdn.net/''~\footnote{Prefix search URL for some Google ads: \url{https://web.archive.org/web/*/https://s0.2mdn.net/*}}, ``https://s-static.innovid.com/''~\footnote{Prefix search URL for Innovid ads: \url{https://web.archive.org/web/*/https://s-static.innovid.com/*}}, and ``https://cdn.flashtalking.com''~\footnote{Prefix search URL for Flashtalking ads: \url{https://web.archive.org/web/*/https://cdn.flashtalking.com*}}.
Perma.cc, Archive Team, and Archive-It users archived or uploaded most of the recent (before August 2023) ads, but Internet Archive prevented Save Page Now users from archiving image, video, and web page ads, as well as the Google SafeFrames and other ad iframes needed to load them. 

\begin{figure}[tbp]
     \centering
     \begin{subfigure}[b]{\textwidth}
        \centering
         \includegraphics[scale=0.35]{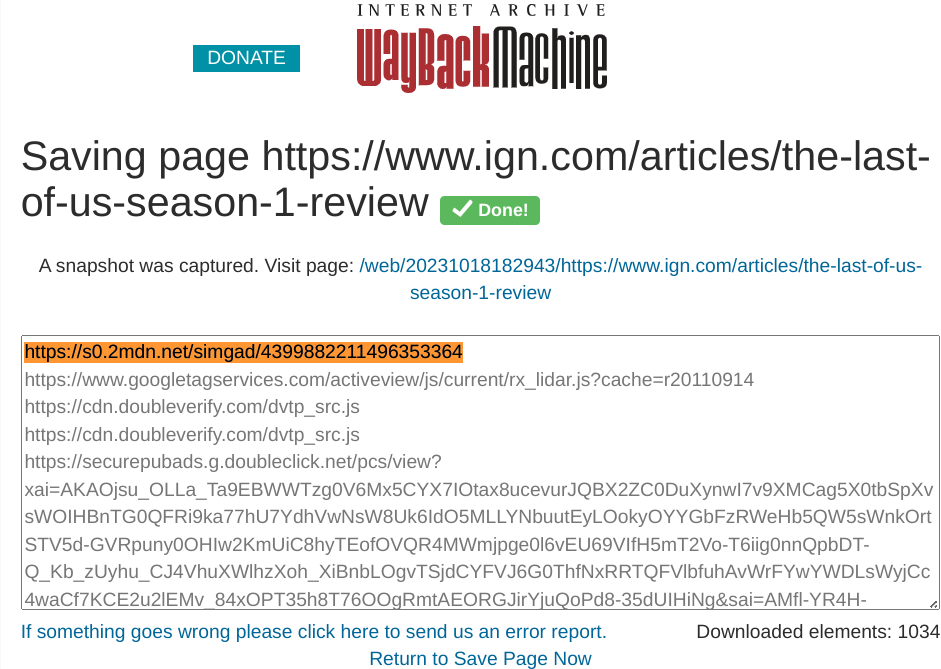}
         \caption{URI-Rs identified during a crawling session. The URI-R for an image ad is highlighted.}
         \label{fig:adIdentifiedBySPN}        
     \end{subfigure}
     \vfill
     \begin{subfigure}[b]{\textwidth}
        \centering
         \includegraphics[scale=0.35]{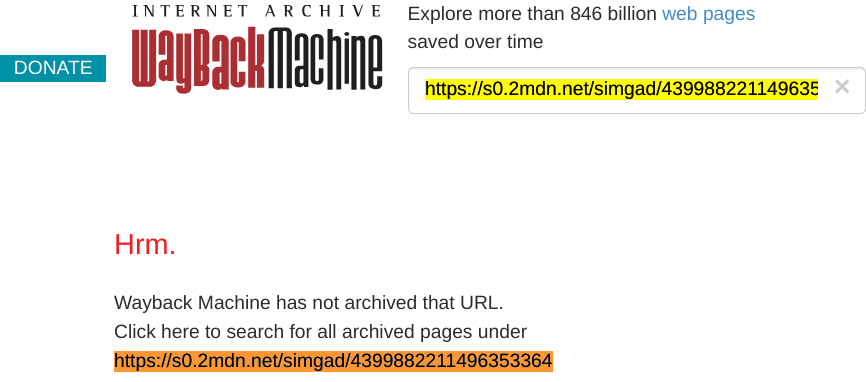}
         \caption{The URI-R for the image ad was not archived.}
         \label{fig:adNotArchivedBySPN}
     \end{subfigure}
     \caption{Save Page Now identified the URI-R for an image associated with an ad during a crawling session, but did not archive the image.}
     \label{fig:adBlockedBySPN}
\end{figure}

After August 2023, Save Page Now allowed the archiving of more ads, provided the user directly archived the URI-R associated with the ad. 
But in October 2023, we archived a web page that loaded Google ads and found that Save Page Now still blocked those ads from being archived (Figure \ref{fig:adBlockedBySPN}). The image ad in Figure \ref{fig:adIdentifiedBySPN} was not archived (Figure \ref{fig:adNotArchivedBySPN}).

\begin{figure}[tbp]
     \centering
     \begin{subfigure}[b]{\textwidth}
        \centering
         \includegraphics[scale=0.3]{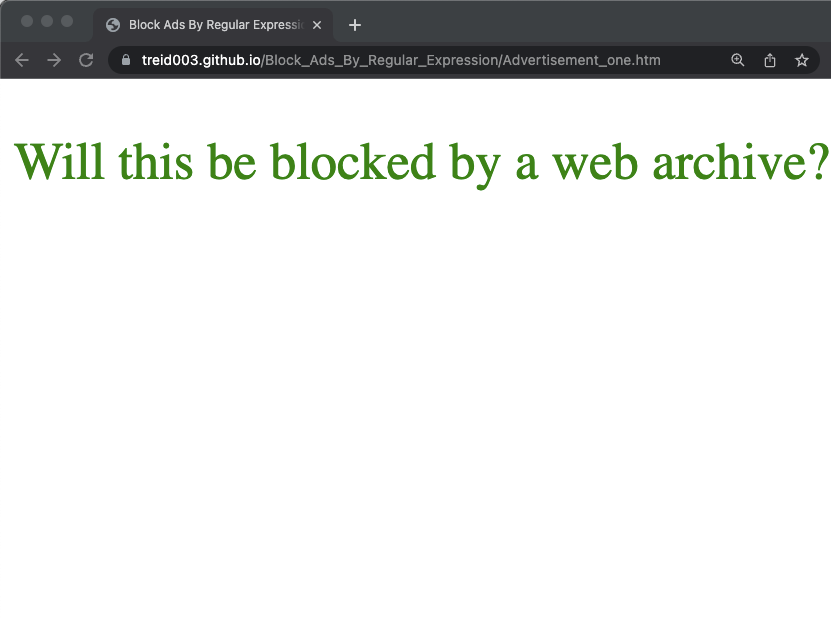}
         \caption{Web page with ``Advertisement'' in the file name. URI-R: \url{https://treid003.github.io/Block_Ads_By_Regular_Expression/Advertisement_one.htm}}
         \label{fig:webPageWithAdRelatedFileName}        
     \end{subfigure}
     \vfill
     \begin{subfigure}[b]{\textwidth}
        \centering
         \includegraphics[scale=0.3]{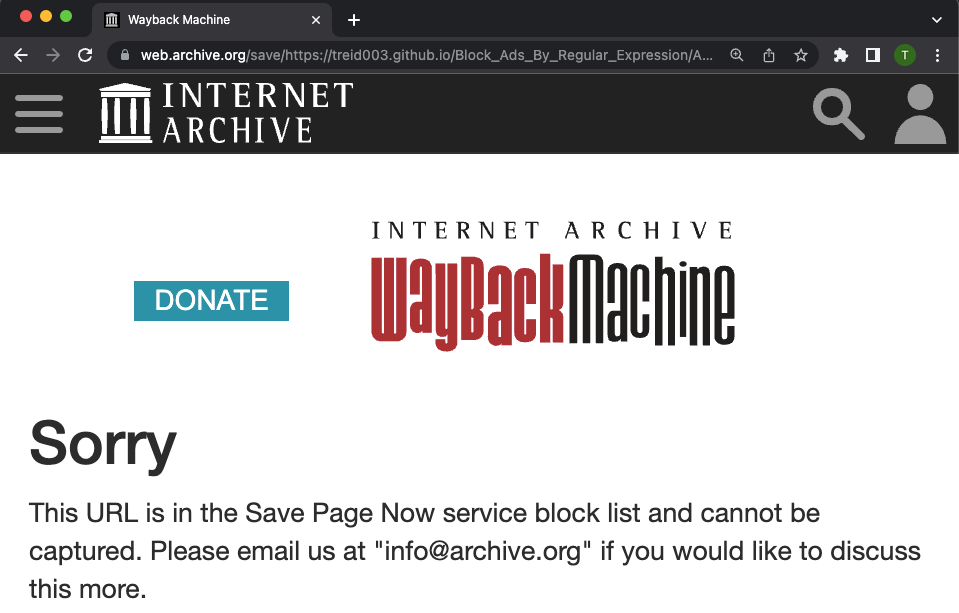}
         \caption{The web page was blocked from being archived}
         \label{fig:SPNblocksWebPageWithAdRelatedName}
     \end{subfigure}
     \caption{Save Page Now previously blocked some web resources from being archived when ad related file names were used}
     \label{fig:webPageBlockedByFileName}
\end{figure}

\begin{figure}[tbp]
     \centering
     \begin{subfigure}[b]{\textwidth}
        \centering
         \includegraphics[width=\textwidth]{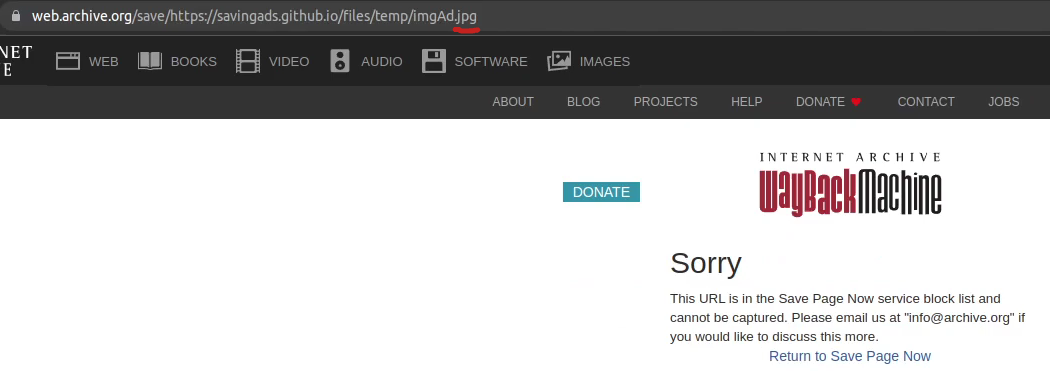}
         \caption{imgAd.jpg was blocked by Save Page Now (video: \url{https://youtu.be/LmGPc7KdcL4?t=264}). This URL is no longer blocked. URI-M: \url{https://web.archive.org/web/20240403212831/https://savingads.github.io/files/temp/imgAd.jpg}}
         \label{fig:imgAdWithExtensionBlocked}        
     \end{subfigure}
     \vfill
     \begin{subfigure}[b]{\textwidth}
        \centering
         \includegraphics[width=\textwidth]{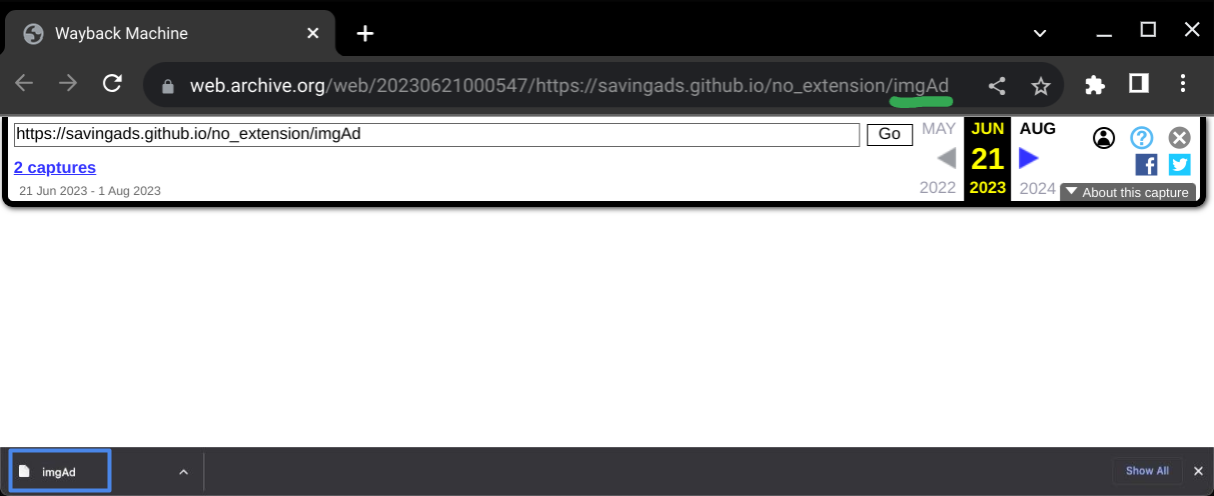}
         \caption{imgAd was not blocked by Save Page Now. URI-M: \url{https://web.archive.org/web/20230621000547/https://savingads.github.io/no_extension/imgAd}}
         \label{fig:imgAdWithoutExtensionArchived}
     \end{subfigure}
     \caption{If the file name does not include a file extension, the ad related file name will not cause the URL to be blocked.}
     \label{fig:imgAdExample}
\end{figure}

\begin{figure}[tbp]
     \centering
     \begin{subfigure}[b]{\textwidth}
        \centering
         \includegraphics[width=\textwidth]{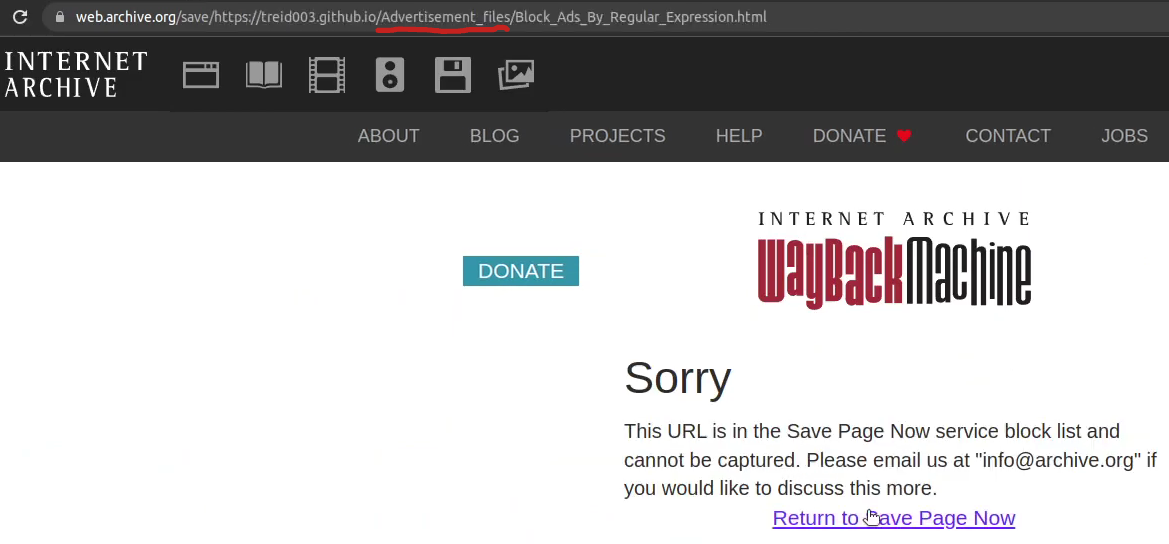}
         \caption{When Block\_Ads\_By\_Regular\_Expression.html was in a directory named Advertisement\_files the URL was blocked by Save Page Now (video: \url{https://youtu.be/MflGE016o28?t=3301}). This URL is no longer blocked. URI-M: \url{https://web.archive.org/web/20240403234751/https://treid003.github.io/Advertisement_files/Block_Ads_By_Regular_Expression.html}} 
         \label{fig:directoryWithAdName}        
     \end{subfigure}
     \vfill
     \begin{subfigure}[b]{\textwidth}
        \centering
         \includegraphics[width=\textwidth]{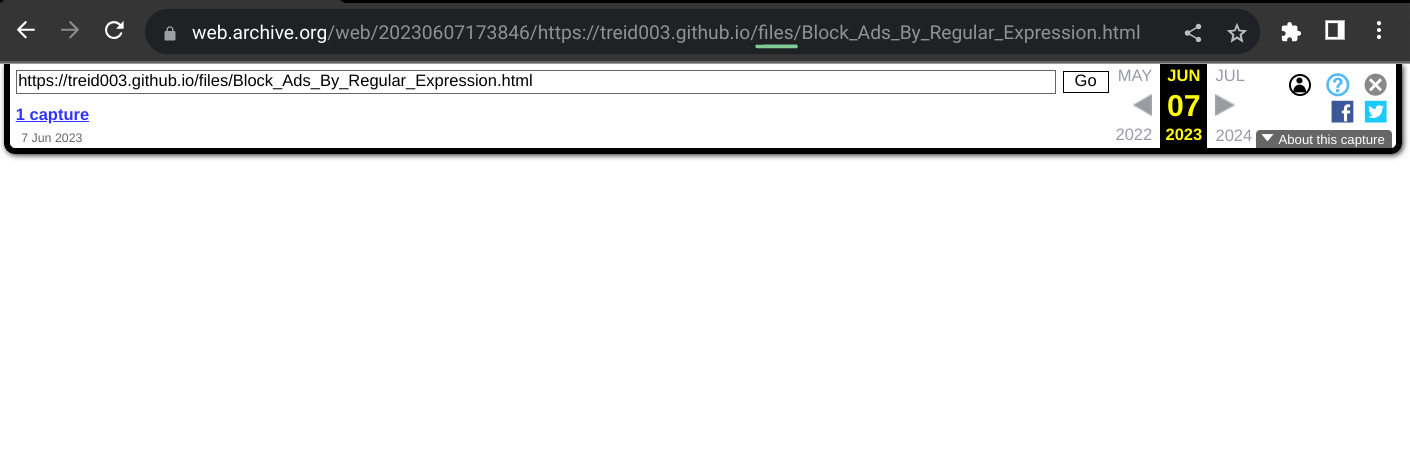}
         \caption{Block\_Ads\_By\_Regular\_Expression.html was archived when the directory it is in did not have an ad related name. URI-M: \url{https://web.archive.org/web/20230607173846/https://treid003.github.io/files/Block_Ads_By_Regular_Expression.html}}
         \label{fig:regularDirectoryName}
     \end{subfigure}
     \caption{When a directory in the URL's path has an ad related name, Save Page Now had blocked the URL.}
     \label{fig:adDirectoryExample}
\end{figure}

Save Page Now also blocked URLs that included an ad-related file or directory name in the URL's path (Figure \ref{fig:webPageBlockedByFileName}). 
To explore these findings further, we created a video playlist~\footnote{Video playlist: \url{https://www.youtube.com/playlist?list=PLYiVfucTlg-MZYrYfLyKF2OR6DbVrDnvf}} to show ad related file names and directory names that caused Save Page Now to block their URLs. 
Blocked file names included ``imgAd.jpg'', ``displayAds.js'', ``videoAd.mp4'', and ``webAd.png''. However, if an ad-related file name did not include a file extension (like a file named ``imgAd''), Save Page Now did not block it (Figure \ref{fig:imgAdExample}).  
Blocked directory names included ``Advertisement\_files'' (Figure \ref{fig:adDirectoryExample}), ``displayAds'', ``videoAd'', ``webAd'', and ``ads''. 

\begin{figure}[tbp]
     \centering
     \begin{subfigure}[b]{\textwidth}
        \centering
         \includegraphics[width=\textwidth]{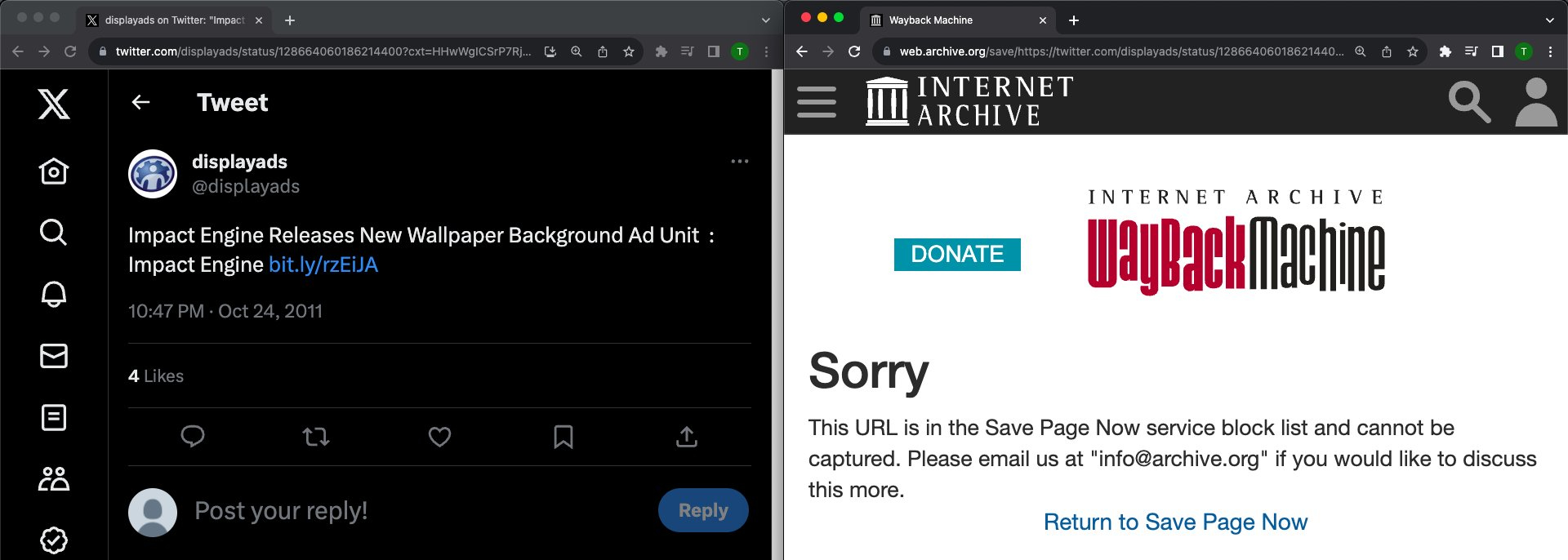}
         \caption{Social media accounts with ad related usernames were getting blocked by Save Page Now.}
         \label{fig:twitterAccountBlockedBySPN}        
     \end{subfigure}
     \vfill
     \begin{subfigure}[b]{\textwidth}
        \centering
         \includegraphics[width=\textwidth]{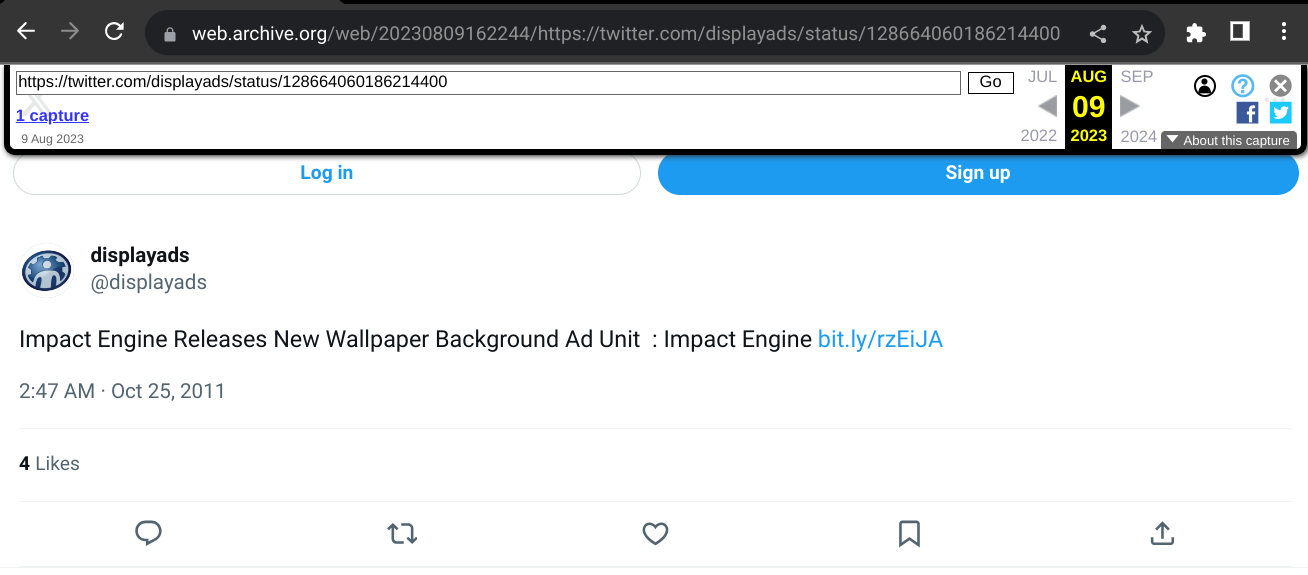}
         \caption{It is now possible to archive social media accounts with an ad related username. URI-M: \url{https://web.archive.org/web/20230809162244/https://twitter.com/displayads/status/128664060186214400}}
         \label{fig:twitterAccountArchivedBySPN}
     \end{subfigure}
     \caption{Before June 2023, some social media accounts with ad related usernames were blocked by Save Page Now}
     \label{fig:socialMediaAccountWithAdUsernames}
\end{figure}

\begin{figure}[tbp]
     \centering
     \begin{subfigure}[b]{0.4\textwidth}
        \centering
         \includegraphics[width=\textwidth]{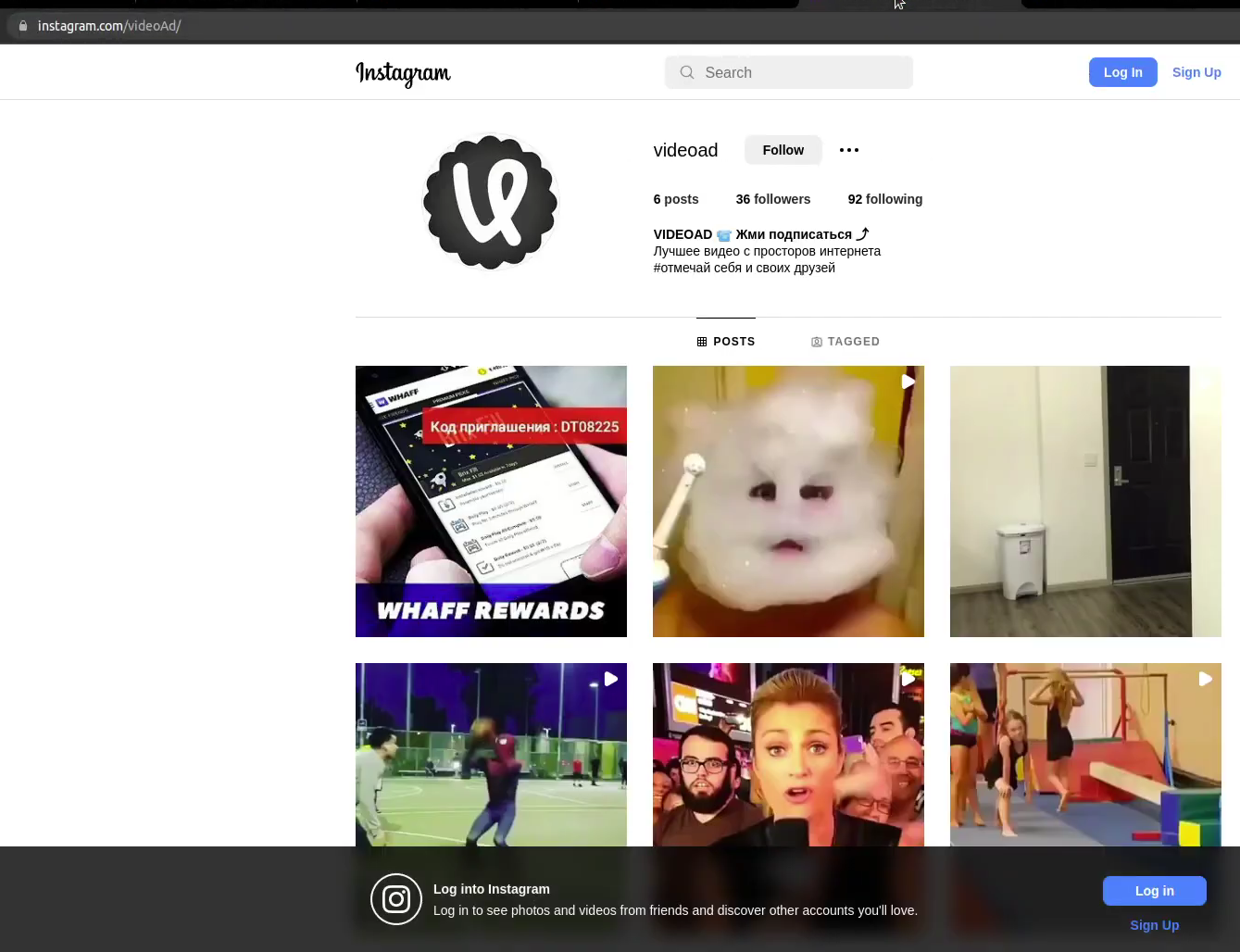}
         \caption{videoad account on Instagram. URI-R: \url{https://www.instagram.com/videoAd/}}
         \label{fig:videoAdInstagramLive}        
     \end{subfigure}
     \hfill
     \begin{subfigure}[b]{0.5\textwidth}
        \centering
         \includegraphics[width=\textwidth, valign=T]{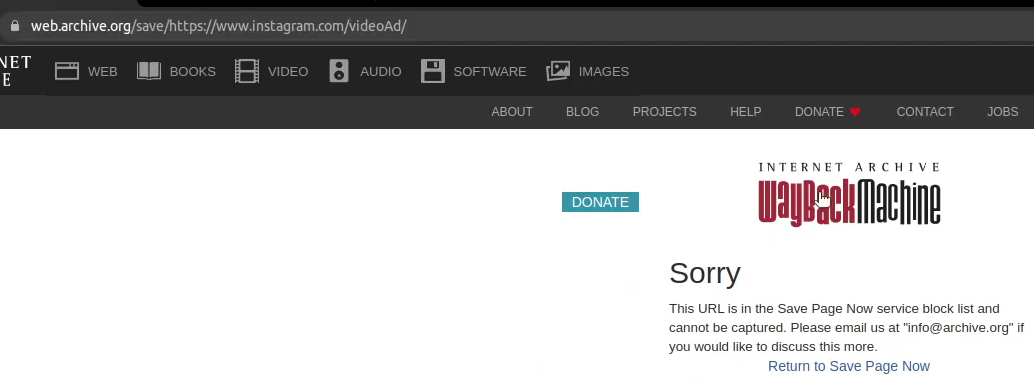}
         \caption{videoad account on Instagram was blocked by Save Page Now. Video: \url{https://youtu.be/vskyvNrdjqw?t=2653}}
         \label{fig:videoAdInstagramBlocked}
     \end{subfigure}
     \vfill
     
     \begin{subfigure}[b]{0.4\textwidth}
        \centering
         \includegraphics[width=\textwidth]{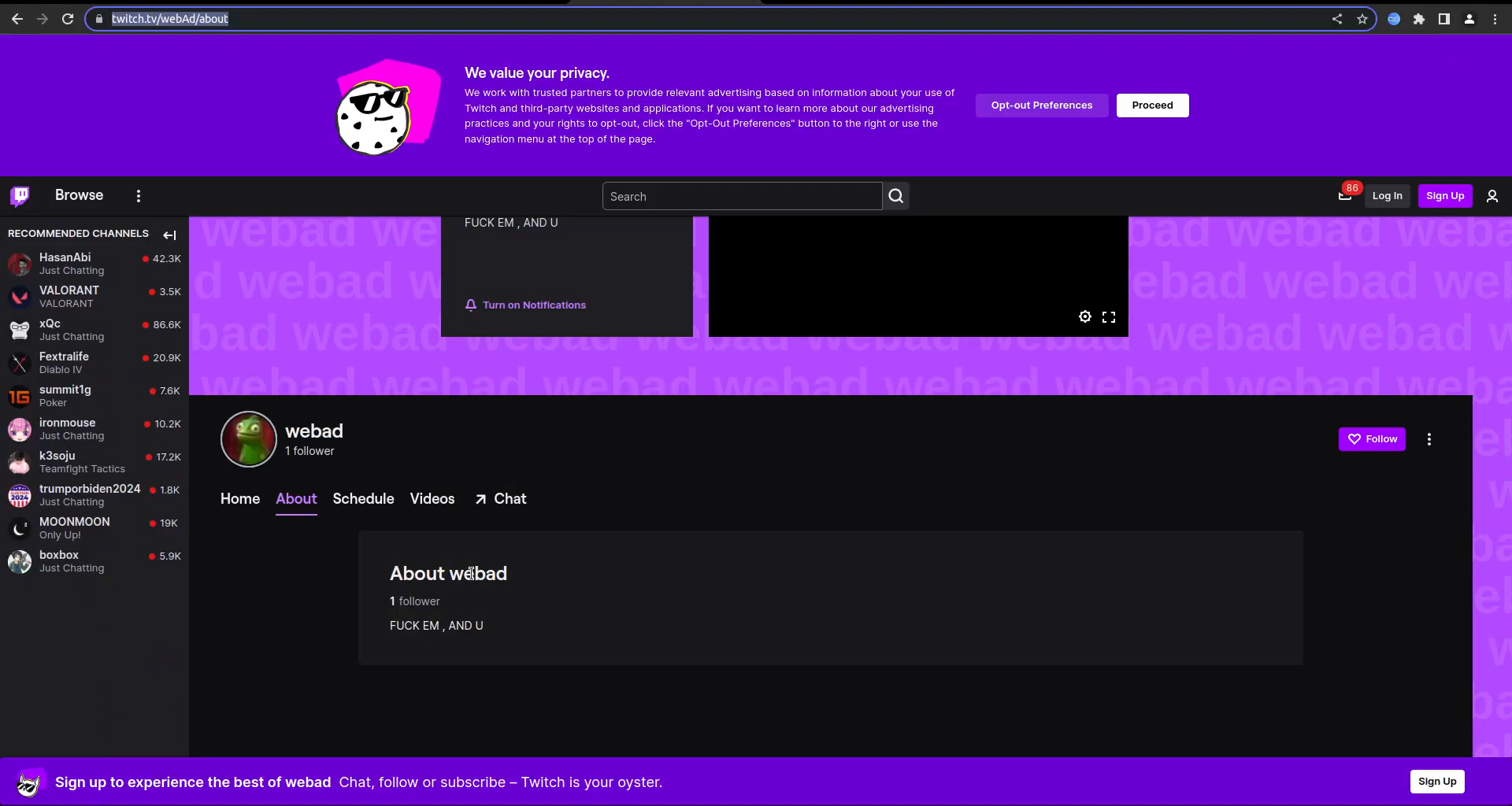}
         \caption{webad account on Twitch. URI-R: \url{https://www.twitch.tv/webAd/about}}
         \label{fig:webAdTwitchLive}        
     \end{subfigure}
     \hfill
     \begin{subfigure}[b]{0.5\textwidth}
        \centering
         \includegraphics[width=\textwidth]{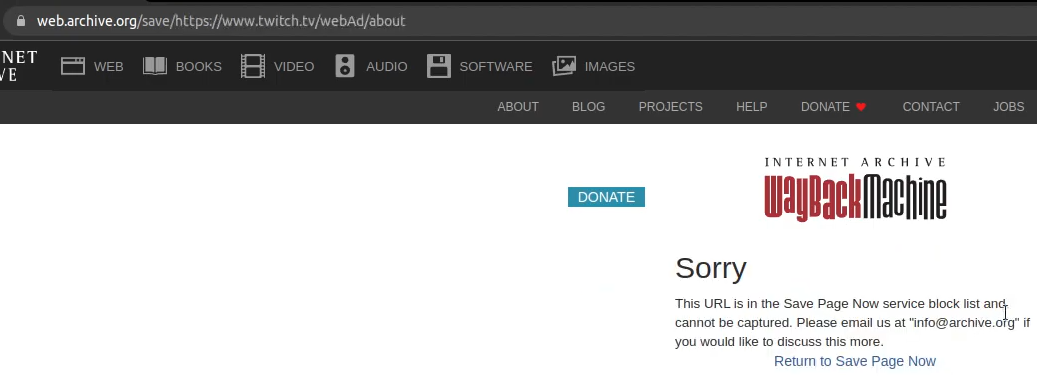}
         \caption{webad account on Twitch was blocked by Save Page Now. Video: \url{https://youtu.be/vskyvNrdjqw?t=3046}}
         \label{fig:webAdTwitchBlocked}
     \end{subfigure}    
      \vfill

     \begin{subfigure}[b]{0.4\textwidth}
        \centering
         \includegraphics[width=\textwidth]{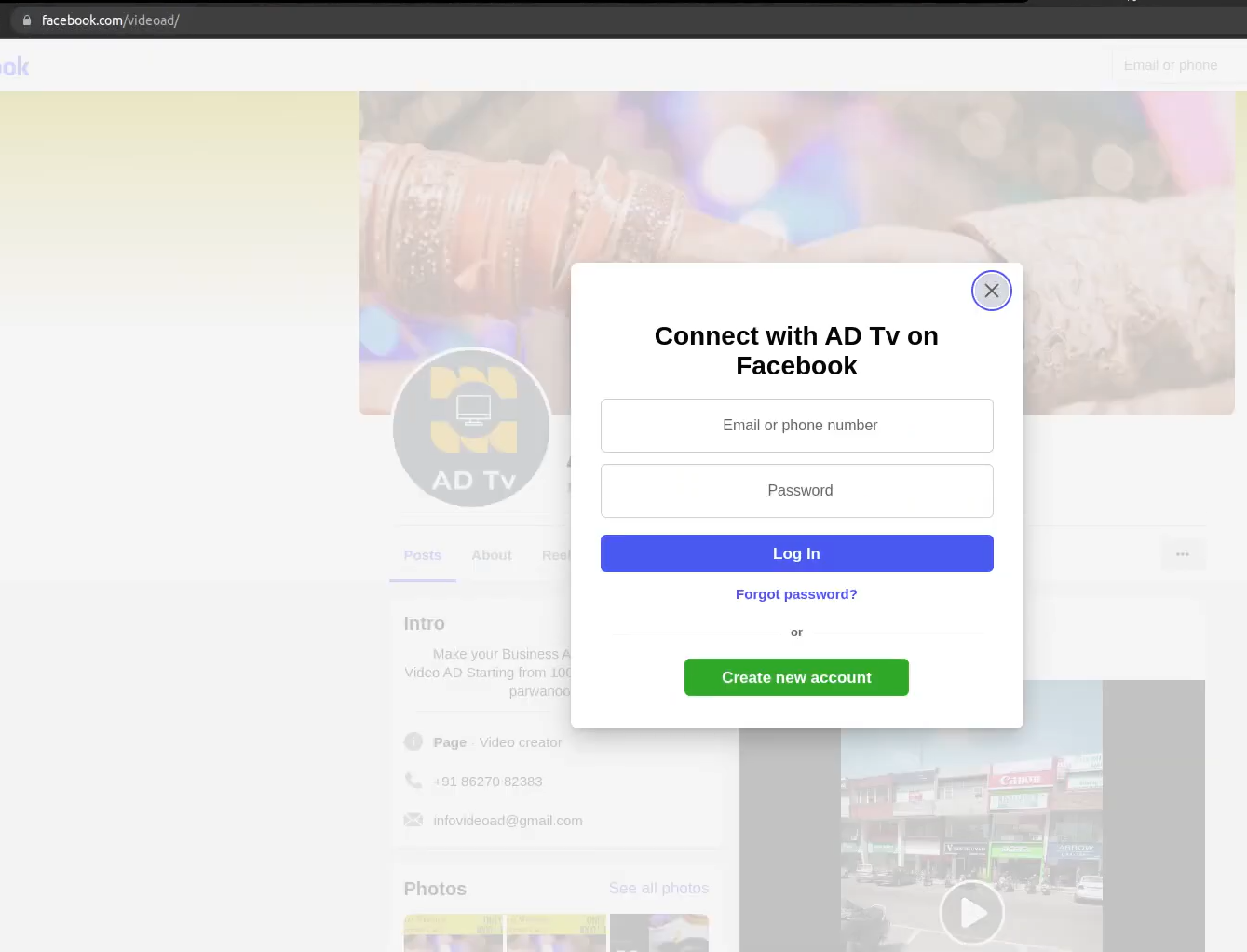}
         \caption{videoad account on Facebook. URI-R: \url{https://www.facebook.com/videoAd/}}
         \label{fig:videoAdFacebookLive}        
     \end{subfigure}
     \hfill
     \begin{subfigure}[b]{0.5\textwidth}
        \centering
         \includegraphics[width=\textwidth]{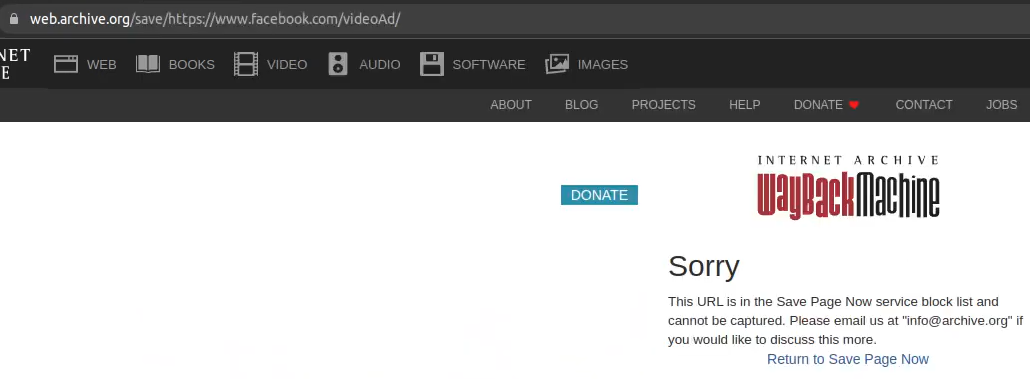}
         \caption{videoad account on Facebook was blocked by Save Page Now. Video: \url{https://youtu.be/vskyvNrdjqw?t=2760}}
         \label{fig:videoAdFacebookBlocked}
     \end{subfigure} 
     \caption{Example of other social media accounts that were previously blocked by Save Page Now}
     \label{fig:otherSocialMediaAccountsBlocked}
\end{figure}

As of June 2023, Save Page Now also blocked social media accounts with an ad-related username if the social media platform used the username as a directory in the URL's path (Figure \ref{fig:twitterAccountBlockedBySPN}). For example, Save Page Now blocked our effort to archive \texttt{displayads}'s tweet (Figure \ref{fig:twitterAccountBlockedBySPN}) because the name ``displayads'' was a directory in the URL's path. We communicated with Wayback Machine staff about this issue (in August of 2023) and they made this account archivable (Figure \ref{fig:twitterAccountArchivedBySPN}). 
Save Page Now also blocked ads on social media websites like Instagram (Figures \ref{fig:videoAdInstagramLive} and \ref{fig:videoAdInstagramBlocked}), Twitch (Figures \ref{fig:webAdTwitchLive} and \ref{fig:webAdTwitchBlocked}), and Facebook (Figures \ref{fig:videoAdFacebookLive} and \ref{fig:videoAdFacebookBlocked}). 

\subsubsection{Brozzler’s Incompatibility With Recent Versions of Google Chrome}
\begin{figure}[tbp]
    \centering
    \includegraphics[scale=0.175]{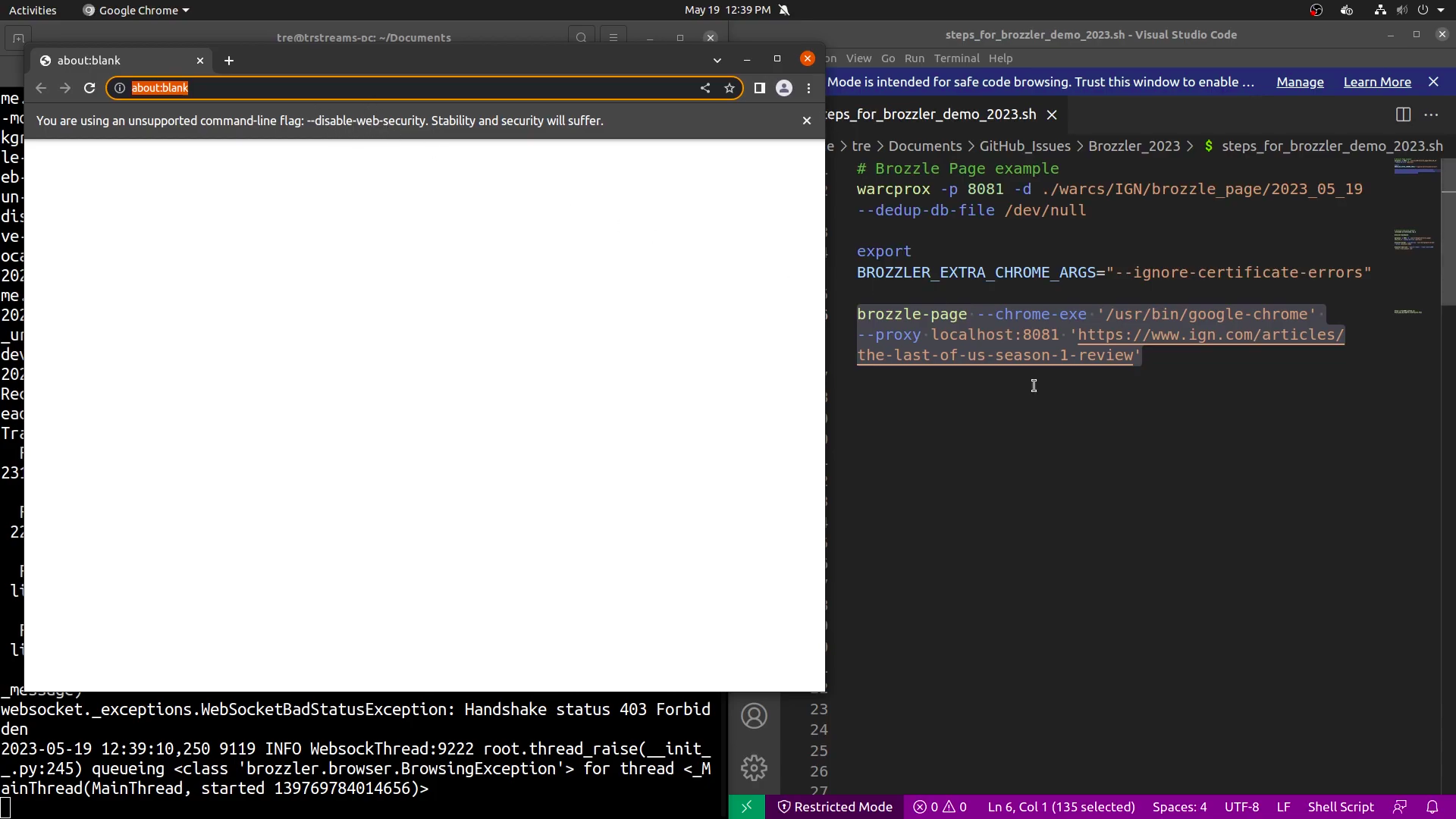}
    \caption{The web page does not load when Brozzler is archiving a web page with a Chrome version released after version 110. GitHub issue: \url{https://github.com/internetarchive/Brozzler/issues/256}}
    \label{fig:webPageFailedToLoad}
\end{figure}

\begin{figure}[tbp]
    \centering
    \includegraphics[scale=0.19]{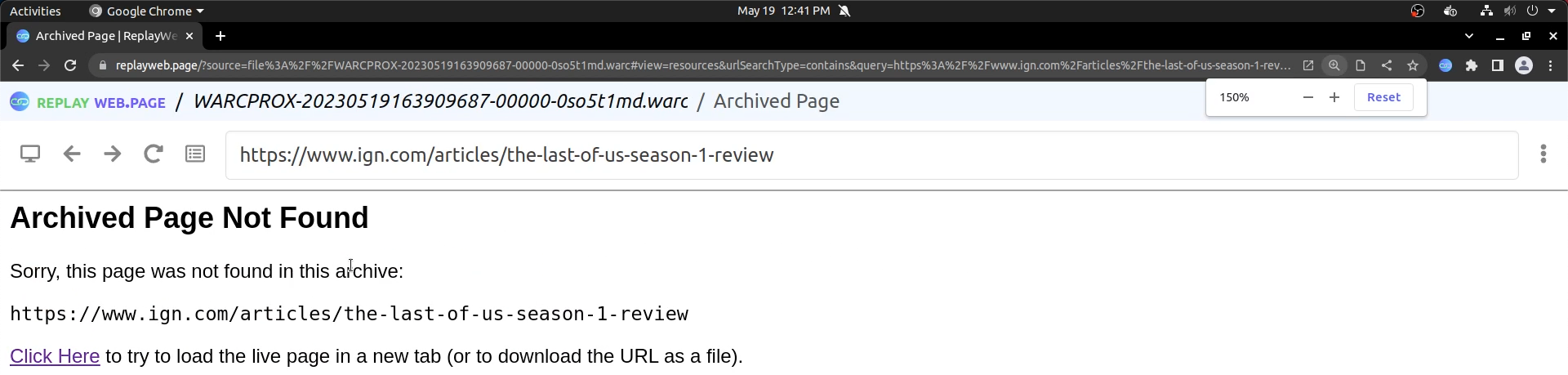}
    \caption{Brozzler failed to archive a web page (https://www.ign.com/articles/the-last-of-us-season-1-review) with Chrome version 113.0.5672.126. WARC file: \url{https://zenodo.org/records/10373135/files/WARCPROX-20230519163909687-00000-0so5t1md.warc?download=1}}
    \label{fig:failedToArchive}
\end{figure}

\begin{figure}[tbp]
    \centering
    \includegraphics[scale=0.20]{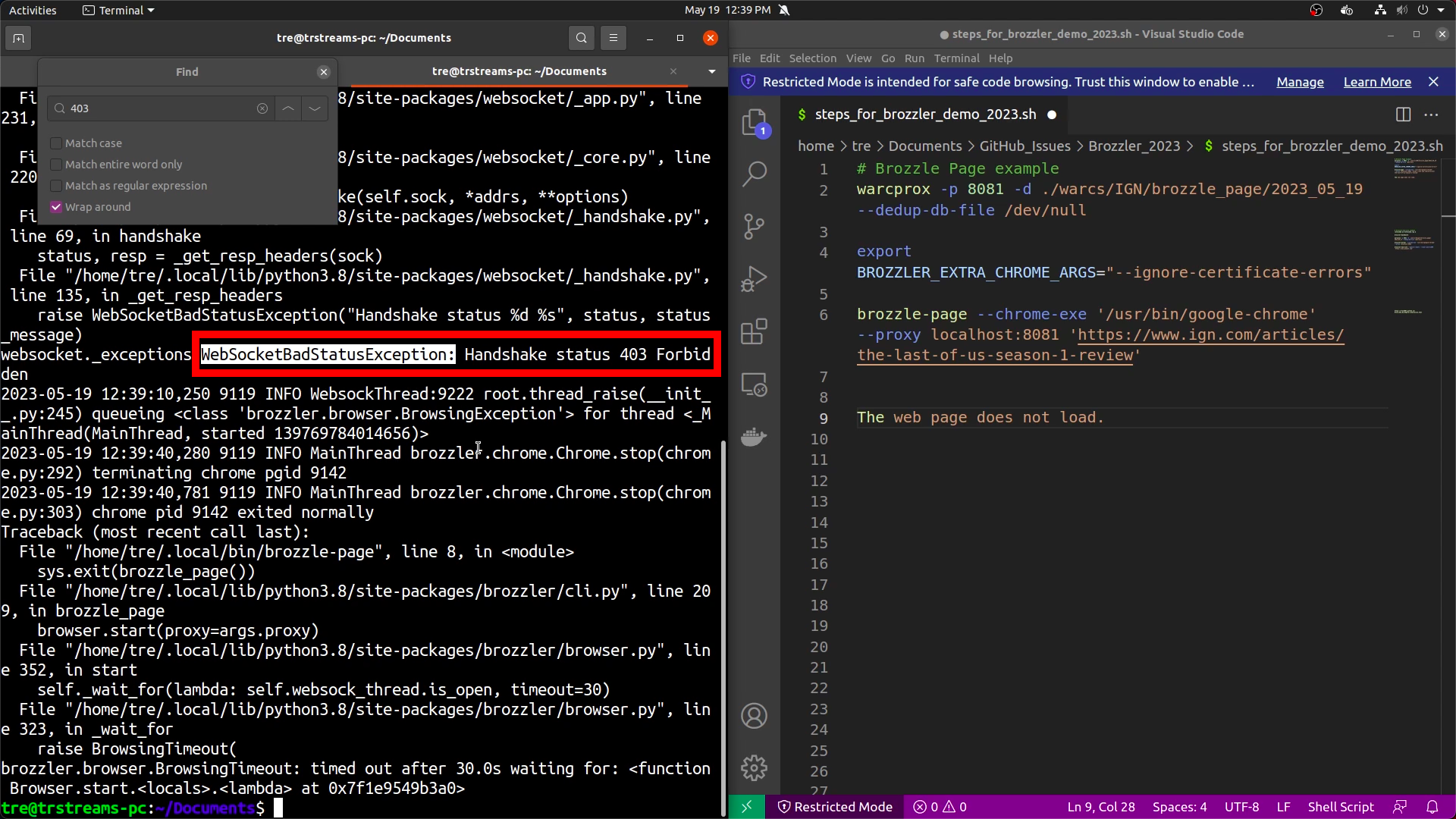}
    \caption{A WebSocketBadStatusException occurs when using Brozzler to archive web pages with versions of Chrome after version 110.}
    \label{fig:webSocketBadStatusException}
\end{figure}

\begin{figure}[tbp]
    \centering
    \includegraphics[scale=0.25]{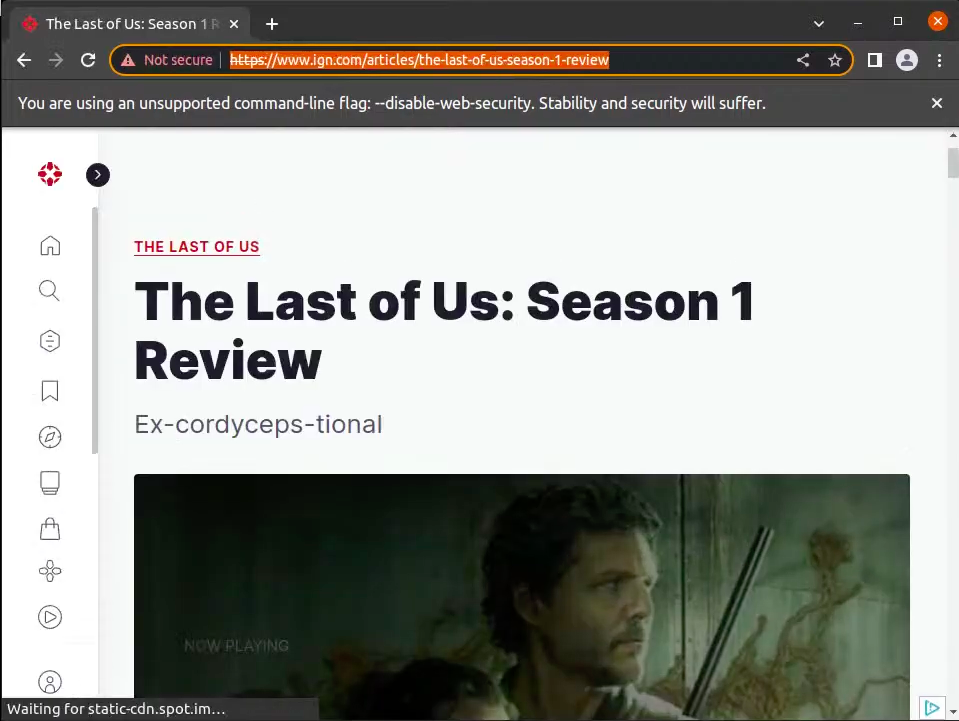}
    \caption{Brozzler was able to load web pages during crawl time with earlier versions of Chrome released at the beginning of 2023. Video: \url{https://www.youtube.com/live/0n_CcYm1Z90?feature=share&t=41} }
    \label{fig:loadedWebPage}
\end{figure}

Brozzler became incompatible~\footnote{\url{https://github.com/internetarchive/brozzler/issues/256}} with versions of Google Chrome that released after March 2023.
When we attempted to archive a web page using the \texttt{brozzle-page} command (Listing \ref{BrozzlerCommands}), Brozzler failed to load the web page (Figure \ref{fig:webPageFailedToLoad}), which prevented it from being archived (Figure \ref{fig:failedToArchive}). 
When we executed these commands on Ubuntu (22.04.2 and 20.04.6 LTS) and macOS (Ventura 13.3.1), a ``WebSocketBadStatusException: Handshake status 403 Forbidden'' error occurred (Figure \ref{fig:webSocketBadStatusException}).  
When we employed these same commands in early 2023, however, Brozzler loaded the desired web page (Figure \ref{fig:loadedWebPage}).  

\noindent\begin{minipage}[tb]{\textwidth} \begin{lstlisting}[language=bash, breaklines=true, label=BrozzlerCommands, caption=The commands that were used for this example. Video: \url{https://youtu.be/A-zr6zVTZSo?t=4888}]
warcprox -p 8081 -d ./warcs/IGN/brozzle_page/2023_05_19 
--dedup-db-file /dev/null
export Brozzler_EXTRA_CHROME_ARGS="--ignore-certificate-errors" 
brozzle-page --chrome-exe '/usr/bin/google-chrome' --proxy localhost:8081 'https://www.ign.com/articles/the-last-of-us-season-1-review' \end{lstlisting} \end{minipage}

\newpage
After testing Brozzler on a range of web pages, we concluded that browser incompatibility is to blame.
The last stable version of Google Chrome that worked with Brozzler is version 110.0.5481.177, (released\footnote{Chrome 110.0.5481.177: \url{https://chromereleases.googleblog.com/2023/02/stable-channel-desktop-update_22.html}} on February 22, 2023). Originally released in mid-December 2023, the Chrome version (120.0.6099.109)\footnote{Chrome 120.0.6099.109: \url{https://chromereleases.googleblog.com/2023/12/stable-channel-update-for-desktop_12.html}} remained incompatible with Brozzler~\footnote{Video: \url{https://youtu.be/6eIa8A87Rq8}}.

\subsection{Replay Problems}
When we sought to replay ads from our dataset, we encountered three obstacles. First, the JavaScript for Google and Amazon ad services dynamically generated URLs with random values. Second, the JavaScript for Flashtalking ad service prevented the replay of an embedded web page ad outside of an ad iframe. Finally, some ads did not load during replay depended on the browser.

\subsubsection{Replaying Google SafeFrames} \label{Replaying_Google_SafeFrame}
\begin{figure}[tbp]
    \centering
    \includegraphics[width=\textwidth]{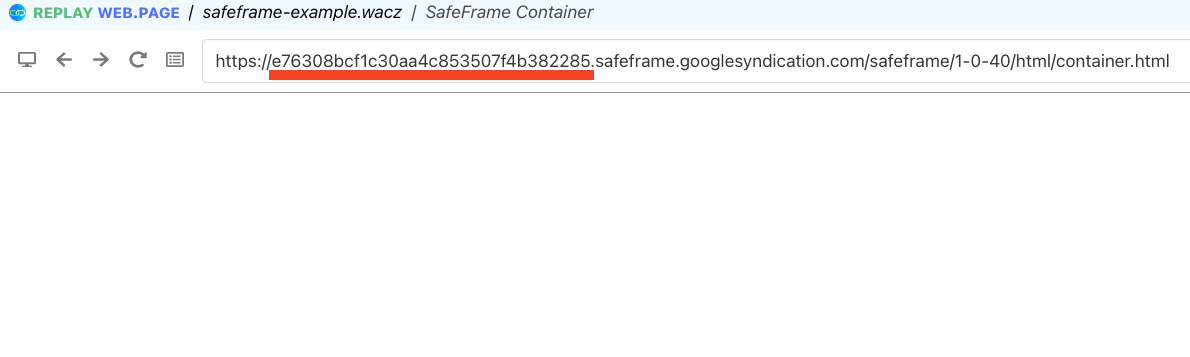}
    \caption{Google SafeFrame successfully archived by ArchiveWeb.page. WACZ: \url{https://zenodo.org/records/10373131/files/safeframe-example.wacz?download=1} | SafeFrame URI-R: \url{https://e76308bcf1c30aa4c853507f4b382285.safeframe.googlesyndication.com/safeframe/1-0-40/html/container.html}}
    \label{fig:safeFrameArchived}
\end{figure}

\begin{figure}[tbp]
    \centering
    \includegraphics[scale=0.2]{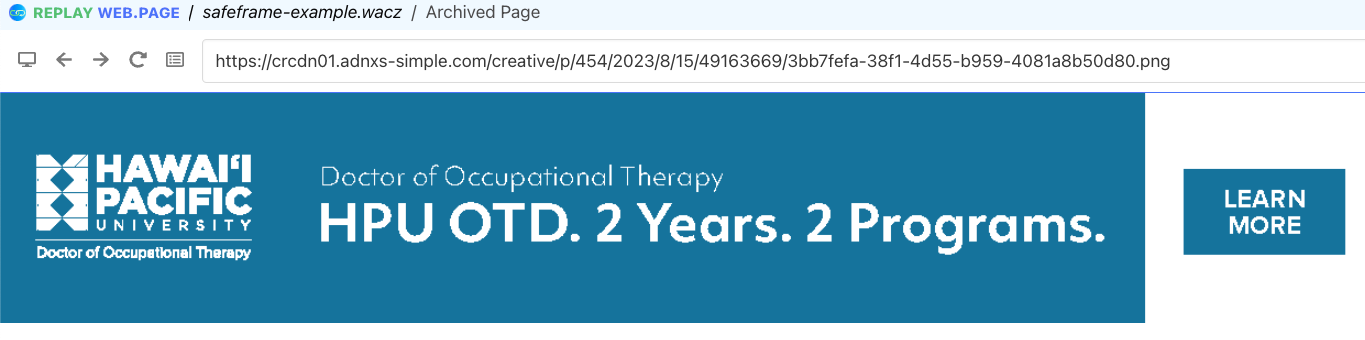}
    \caption{Ad content successfully archived by ArchiveWeb.page}
    \label{fig:safeFrameAdArchived}
\end{figure}

Google SafeFrame uses a random value in the iframe's URL's subdomain (Figure \ref{fig:safeFrameURI}), which prevents replay.
\begin{table}[tbp]
\begin{tabular}{|l|l|}
\hline
\multicolumn{1}{|c|}{\begin{tabular}[c]{@{}c@{}}Replay Session \\ Number\end{tabular}} & \multicolumn{1}{c|}{Random Value in Google SafeFrame URL} \\ \hline
1                                                                                      & af393d3d232450caab92d97eaefb484e                \\ \hline
2                                                                                      & 36dc52191b8e81186b187c938af4b280                \\ \hline
3                                                                                      & 5f68c90c97e25bf663f52ef786eb49b8                \\ \hline
4                                                                                      & f663f52ef786eb49b8d803369fd0abea                \\ \hline
5                                                                                      & 6d18ef14a03123734d453dee25a8be6e                \\ \hline
6                                                                                      & 7a9317739c43b98082f4e77ed17fe3fa                \\ \hline
7                                                                                      & 4f8332dc6d18ef14a03123734d453dee                \\ \hline
8                                                                                      & 5bf663f52ef786eb49b8d803369fd0ab                \\ \hline
9                                                                                      & 227cd10c62e4b3ea26c2bd42e587e2c5                \\ \hline
10                                                                                     & 173e9e9bad424f8332dc6d18ef14a031                \\ \hline
\end{tabular}
\caption{When loading the ad iframe (Google SafeFrame) for a Google ad, the random value in the Google SafeFrame URL differed each time the archived web page was replayed. (URI-R: \url{https://mortalkombat.fandom.com/wiki/Tag_Team_Ladder} | WARC: \url{https://zenodo.org/record/7601187/files/2023-01-11_00-59-34_ads_on_fandom_browsertrix_crawler.warc.gz?download=1})} \label{table:SafeFrameSubdomainChangesWhenLoadingAdDuringReplay}
\end{table}
When the seven replay systems we tested (pywb, OpenWayback, ReplayWeb.page, Conifer, Wayback Machine, and Arquivo.pt) executed JavaScript code that generated a random number, the random value differed from the random value generated during crawl time. These replay systems also generated different random values on each replay. Drawing upon public web archives, Aturban et al. \cite{aturban-plosone-2023} discussed this problem of inconsistent replay of archived web pages.
\begin{figure}[tbp]
    \centering
    \includegraphics[scale=0.20]{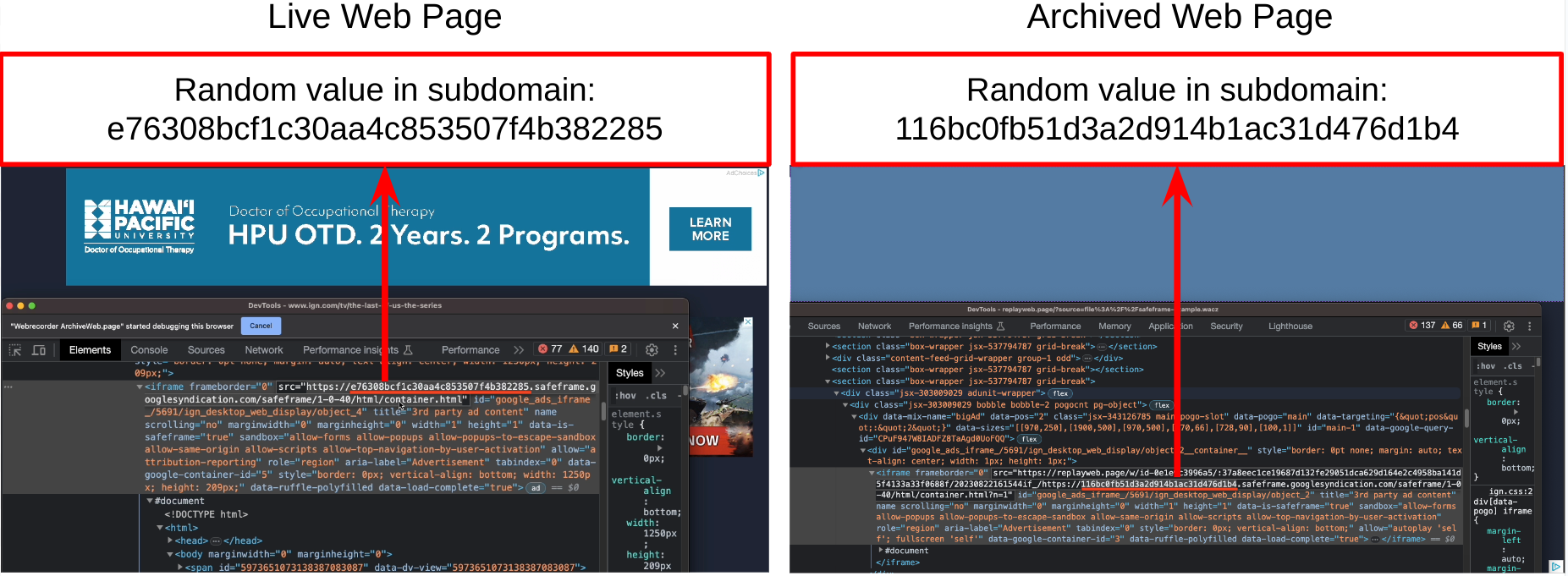}
    \caption{Currently web archive replay systems cannot generate the same random subdomain that was used during crawl time}
    \label{fig:safeFrameLivevsReplay}
\end{figure}
Since we successfully archived the Google SafeFrame and the ad content (Figures \ref{fig:safeFrameArchived} and \ref{fig:safeFrameAdArchived}), this problem was caused by the replay system, not the crawler. 
When ReplayWeb.page, pywb, Conifer, or Wayback Machine attempted to load a Google SafeFrame for an advertisement, an HTTP status code of 404 (Not Found) was returned. 
In contrast, when loading a Google SafeFrame with Arquivo.pt, an HTTP status code of 307 (Temporary Redirect) was returned. Arquivo.pt's replay system changed the timestamp for the archived Safeframe's URI-M, but failed to load the ad content into the successfully archived SafeFrame. 


\begin{figure}[tbp]
    \centering
    \includegraphics[width=0.9\textwidth]{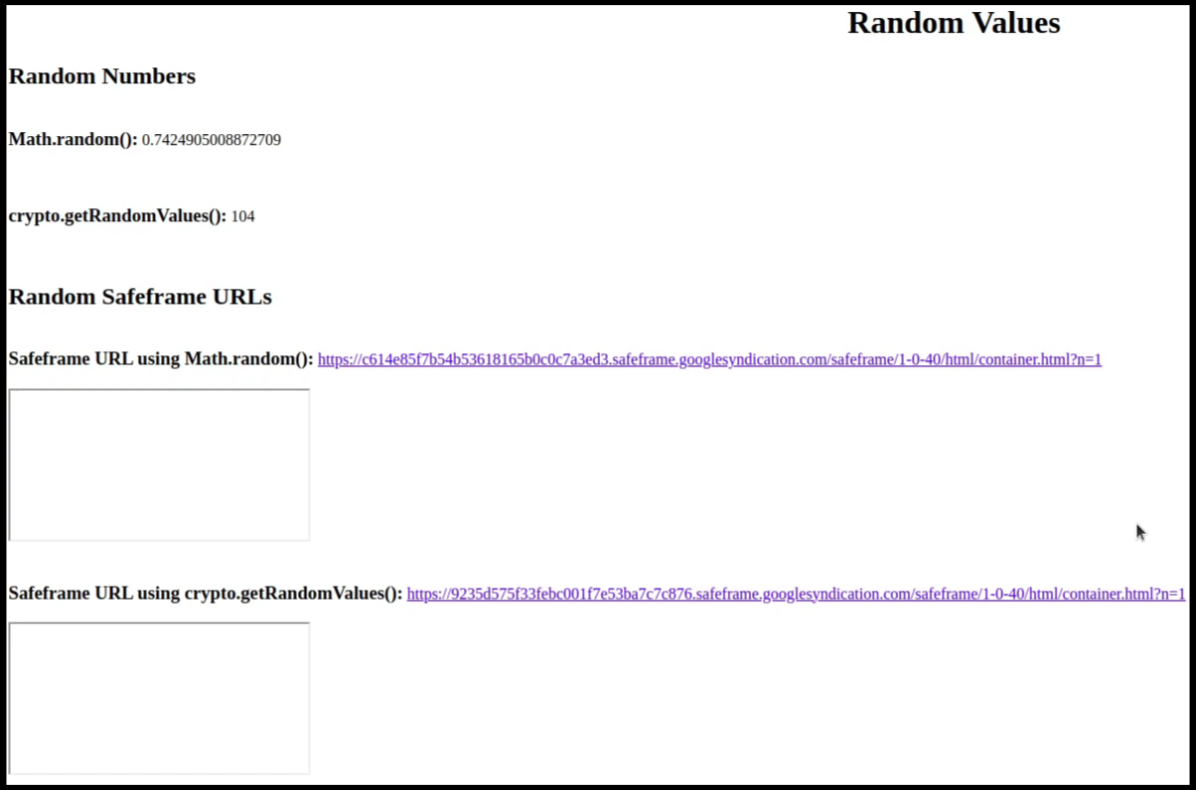}
    \caption{Demo web page (shown during the crawling session) that generates random values and Google SafeFrames. URI-R: \url{https://treid003.github.io/random_Values_external_JS_with_async.html}}
    \label{fig:randomValuesAndSafeFrameCrawlTime}
\end{figure}

Since the random value in a Google SafeFrame URL changed each time a Google ad was replayed (Table \ref{table:SafeFrameSubdomainChangesWhenLoadingAdDuringReplay}), we created an example web page \footnote{Demo web page for generating random numbers and Google SafeFrames: \url{https://treid003.github.io/random_Values_external_JS_with_async.html}} (Figure \ref{fig:randomValuesAndSafeFrameCrawlTime}) that used ad code from Google's \href{https://web.archive.org/web/20230113005605id_/https://securepubads.g.doubleclick.net/gpt/pubads_impl_2023010901.js?cb=31071543}{pubads\_impl\_2023020201.js}~\footnote{URI-M: \url{https://web.archive.org/web/20230113005605id_/https://securepubads.g.doubleclick.net/gpt/pubads_impl_2023010901.js?cb=31071543}} script to determine how the random values were generated for a Google SafeFrame. 
First, we generated random numbers using the same random functions as those used for Google SafeFrame (\texttt{Math.random()} \cite{math-random-mozillat24} and \texttt{window.crypto.getRandomValues()} \cite{crypto-random-mozillat24} functions). This permitted us to determine not only if the replay system generated the same random numbers as those generated during crawl time, but if the random numbers differed on each replay. 
To check the random values generated during crawl time, we used ArchiveWeb.page and recorded a video of the crawling session~\footnote{Crawling session: \url{https://youtu.be/IzGMVmLyYGQ?t=2694}}.
Second, we used the code from Google's pubads\_impl\_2023020201.js script to generate the random value for each SafeFrame URL. In the first section of the web page, two SafeFrame URLs were generated: the first URL used \texttt{Math.random()} and the other used \texttt{window.crypto.getRandomValues()}.
As in the first step, we compared the values generated during crawl and replay time.
Lastly, to check if the replay system successfully replayed the archived Google SafeFrames, we created and embedded iframes into the web page for each SafeFrame URL~\footnote{When a Google SafeFrame is loaded without an ad it will be a blank iframe with no content inside of the iframe.}.

\begin{table}[tbp]
\begin{tabular}{lll}
\multicolumn{3}{c}{Random Values in Google SafeFrame URLs}                                                                                                            \\ \cline{2-3} 
\multicolumn{1}{l|}{}             & \multicolumn{1}{c|}{Generated Using Math.random()}    & \multicolumn{1}{c|}{Generated Using crypto.getRandomValues()} \\ \hline
\multicolumn{1}{|l|}{Crawl Time}  & \multicolumn{1}{l|}{c614e85f7b54b53618165b0c0c7a3ed3} & \multicolumn{1}{l|}{9235d575f33febc001f7e53ba7c7c876}         \\ \hline
\multicolumn{1}{|l|}{Replay Time} & \multicolumn{1}{l|}{328a37205b691401e879e520ba757b42} & \multicolumn{1}{l|}{70b75c8cd9262cb991fc0248a75df5d6}         \\ \hline
\end{tabular}
\caption{Random values in Google SafeFrame URLs generated during crawl time and replay time. Video: \url{https://youtu.be/IzGMVmLyYGQ?t=2697}} \label{table:SafeFrameSubdomainsCrawlTimeVsReplayTime}
\end{table}

\begin{figure}[tbp]
    \centering
    \includegraphics[width=\textwidth]{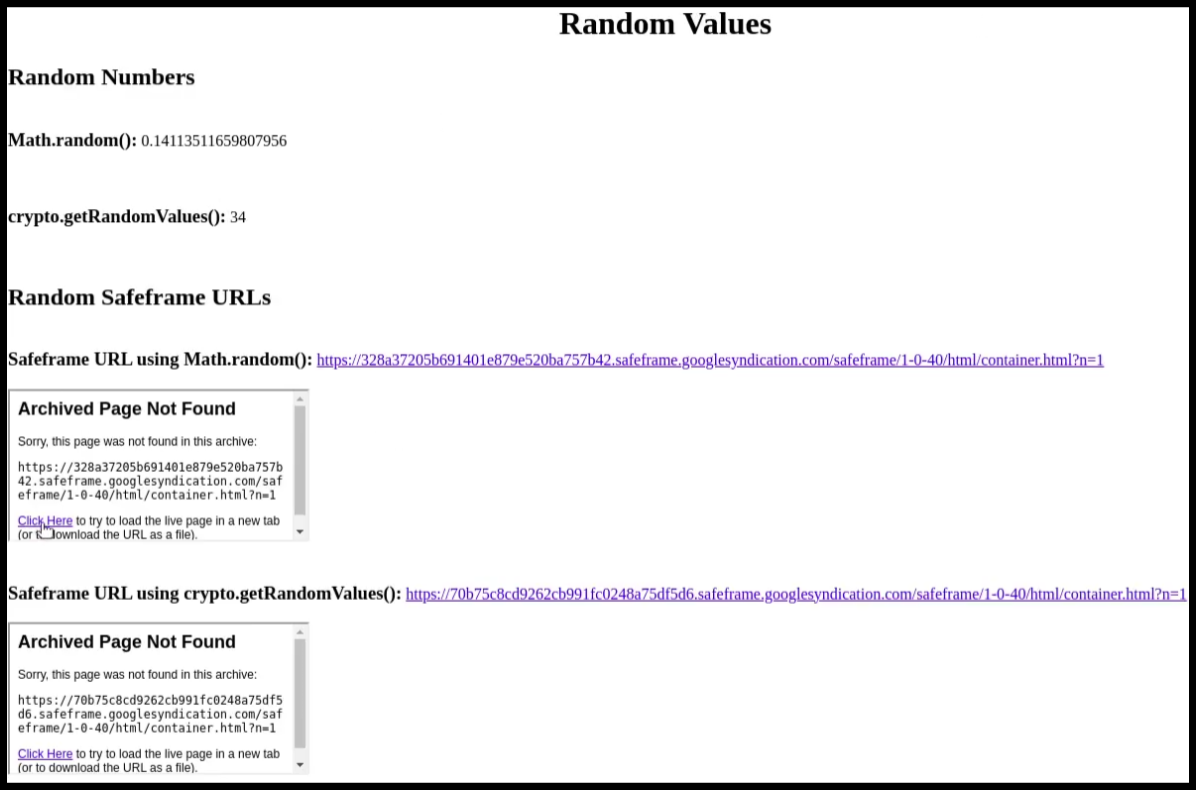}
    \caption{Demo web page shown during the replay session. The values generated during replay differ from the values during crawl time (Figure \ref{fig:randomValuesAndSafeFrameCrawlTime}). WACZ: \url{https://zenodo.org/records/10932857/files/2023-02-21-random-values-with-safeframes.wacz?download=1} | URI-R: \url{https://treid003.github.io/random_Values_external_JS_with_async.html}}
    \label{fig:randomValuesAndSafeFrameReplayTime}
\end{figure}

Table \ref{table:SafeFrameSubdomainsCrawlTimeVsReplayTime} shows the random values generated for the Google SafeFrame URLs during crawl and replay sessions. 
The random values generated during replay and crawl time differed (Figure \ref{fig:randomValuesAndSafeFrameReplayTime}), because the seeds used during the crawling session for the random number generators differed from the seeds used by ReplayWeb.page. Replay systems employ rewriting tools like Wombat.js to overwrite the seed for the random number generators. This results in a more consistent replay where the random values generated should be the same upon each replay.

Listings \ref{wombatMathRandom} and \ref{wombatCryptoRandom} show the JavaScript code used to initialize the \texttt{Math.random()} and \texttt{crypto.getRandomValues()}. When initializing \texttt{Math.random()} (Listing \ref{wombatMathRandom}), Wombat.js overwrote the random number generator's seed (on line 7) with an expression that includes the time of the resource's archiving. Wombat.js initialized \texttt{crypto.getRandomValues()} (Listing \ref{wombatCryptoRandom}) by overwriting the function (lines 6-10).
Seed selection will differ during the crawling and replay sessions because of different implementations for \texttt{Math.random()} and \texttt{crypto.getRandomValues()}.
For \texttt{Math.random()}, the seed selection is an ``implementation-defined'' strategy~\footnote{\url{https://tc39.es/ecma262/multipage/overview.html}}. This allows an external source to define its approach without recommendations from the standard specification \cite{math-random-ecma24}. The W3C API specification \cite{crypto-interface-w3c24} states that the random number generator for \texttt{crypto.getRandomValues()} should be seeded with a high-quality entropy source like ``/dev/urandom'', an operating system entropy source that retrieves environmental noise from device drivers \cite{random-linux-man24}.

\noindent\begin{minipage}[tbp]{\textwidth}
\begin{lstlisting}[language=JavaScript, breaklines=true, label=wombatMathRandom, caption=Wombat.js overriding the seed for Math.random()]
//From Wombat.js. URI-R: https://replayweb.page/static/wombat.js
//Initializing Math.random() by overwriting the seed
f.prototype.initSeededRandom = function(t) {
    this.$wbwindow.Math.seed = parseInt(t);
    var e = this;
    this.$wbwindow.Math.random = function() {
        return e.$wbwindow.Math.seed = (9301 * e.$wbwindow.Math.seed + 49297) % 233280,
        e.$wbwindow.Math.seed / 233280
    }
}

/*...*/

f.prototype.wombatInit = function() { /*...*/ 
    this.initSeededRandom(this.wb_info.wombat_sec) 
    /*...*/}

/*...*/

function f(e, i) {/*...*/ this.wb_info = i, /*...*/}

/*...*/

const b = f; /*...*/   
window._WBWombatInit = function(t) {/*...*/ var e = new b(this,t); /*...*/}

//From the archived web page. URI-R: https://www.ign.com/tv/the-last-of-us-the-series
if (window && window._WBWombatInit) {window._WBWombatInit(wbinfo);}
/*...*/
//wombat_sec is the time when the web page was archived 2023-08-22
wbinfo.wombat_sec = "1692720944";
\end{lstlisting}
\end{minipage}

\noindent\begin{minipage}[t]{\textwidth}
\begin{lstlisting}[language=JavaScript, breaklines=true, label=wombatCryptoRandom, caption=Wombat.js overriding the crypto.getRandomValues() function]
//From Wombat.js. URI-R: https://replayweb.page/static/wombat.js
//Initializing crypto.getRandomValues() by overwriting the function
f.prototype.initCryptoRandom = function() {
    if (this.$wbwindow.crypto && this.$wbwindow.Crypto) {
        var t = this
          , e = function(e) {
            for (var r = 0; r < e.length; r++)
                e[r] = parseInt(4294967296 * t.$wbwindow.Math.random());
            return e
        };
        this.$wbwindow.Crypto.prototype.getRandomValues = e,
        this.$wbwindow.crypto.getRandomValues = e
    }
}
\end{lstlisting}
\end{minipage}

\begin{figure}[tbp]
    \centering
     \begin{subfigure}[b]{0.45\textwidth}
        \centering
         \includegraphics[width=\textwidth]{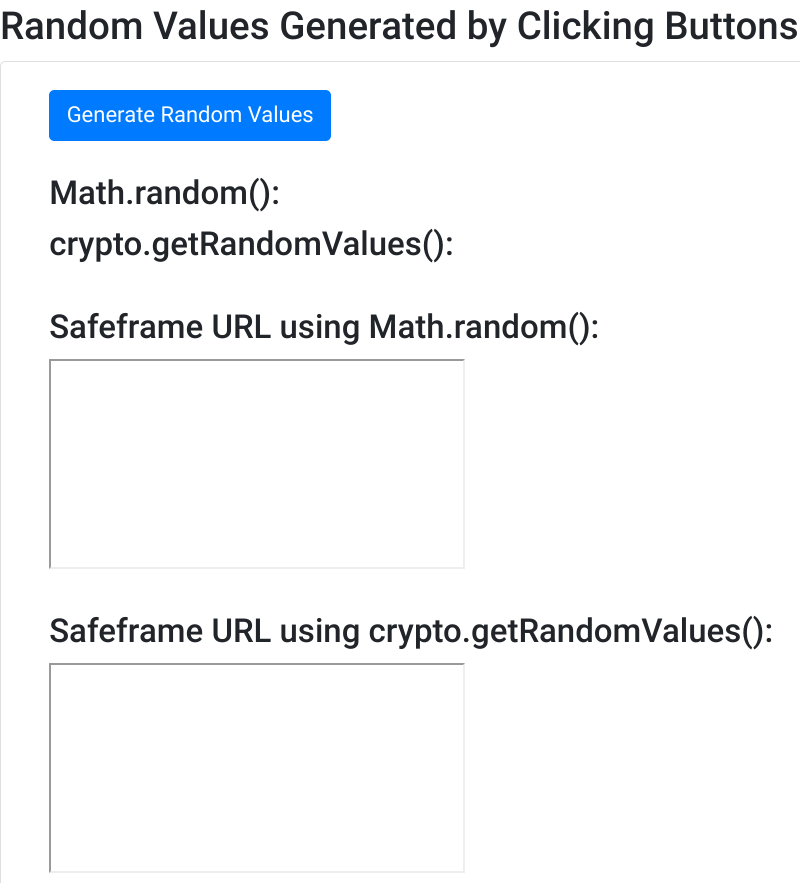}
         \caption{Random values are generated after clicking a button}
         \label{fig:rvButtonClick}        
     \end{subfigure}
     \hfill
     \begin{subfigure}[b]{0.45\textwidth}
        \centering
         \includegraphics[width=\textwidth]{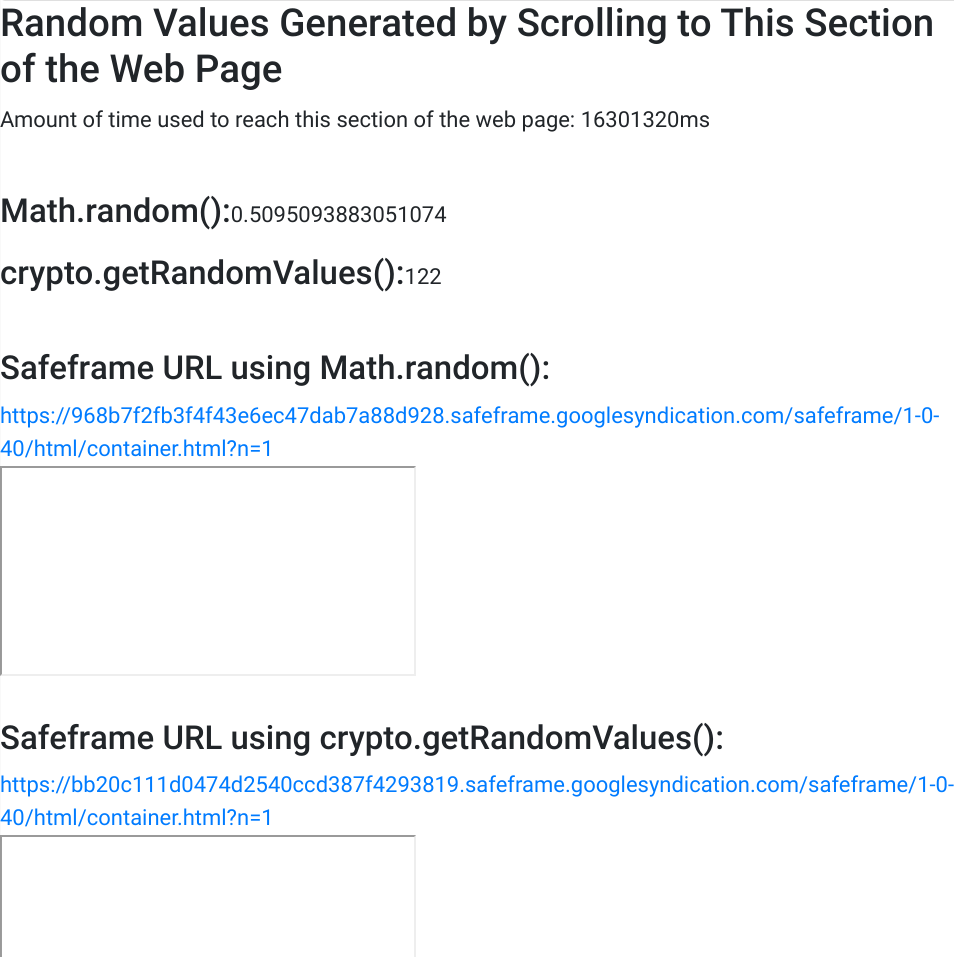}
         \caption{Random values generated after scrolling to this part of the web page}
         \label{fig:rvScrolling}
     \end{subfigure}
    \caption{New sections added to the demo web page that requires user interaction to generate the random values.}
    \label{fig:updatedGSFDemo}
\end{figure}

\begin{table}[tbp]
\begin{tabular}{lll}
                                                                                                                                   & \multicolumn{2}{l}{\begin{tabular}[c]{@{}l@{}}Random Values in Google SafeFrame URLs During Replay for the \\Section  Updated By Scrolling to the Bottom of the Web Page\end{tabular}}                   \\ \hline

\multicolumn{1}{|c|}{\begin{tabular}[c]{@{}c@{}}Total Buttons \\ Clicked Before \\ Scrolling to the \\ last section\end{tabular}} & \multicolumn{1}{c|}{\begin{tabular}[c]{@{}c@{}}Generated Using \\ Math.random()\end{tabular}} & \multicolumn{1}{c|}{\begin{tabular}[c]{@{}c@{}}Generated Using\\ crypto.getRandomValues()\end{tabular}} \\ \hline
\multicolumn{1}{|l|}{0}                                                                                                            & \multicolumn{1}{l|}{75edc22fa0ea2f68a1e0ae51aedf6203}                                         & \multicolumn{1}{l|}{aa89e23772c94a19508c2923a84bed49}                                                   \\ \hline
\multicolumn{1}{|l|}{1}                                                                                                            & \multicolumn{1}{l|}{cfb20651dc8e7d295dbd03dc7dd57827}                                         & \multicolumn{1}{l|}{4f693309e7fad65b456071c053d21559}                                                   \\ \hline
\multicolumn{1}{|l|}{2}                                                                                                            & \multicolumn{1}{l|}{bfc665544160a197051235bd1083f606}                                         & \multicolumn{1}{l|}{c941958a67339666e399eb523c9e282c}                                                   \\ \hline
\multicolumn{1}{|l|}{3}                                                                                                            & \multicolumn{1}{l|}{0772e7dea6a02819114c37a4b768c6d3}                                         & \multicolumn{1}{l|}{b86e07e787554508f07e9c813beeb327}                                                   \\ \hline
\multicolumn{1}{|l|}{4}                                                                                                            & \multicolumn{1}{l|}{0c2dfbe3bba19197b87d2cab13e2e4be}                                         & \multicolumn{1}{l|}{e8b3cf26ef3723cbce88f440ff133635}                                                   \\ \hline
\end{tabular}
\caption{This table shows that the random value in a Google SafeFrame URL can change on each replay if the number of function calls to the \texttt{Math.random()} and \texttt{crypto.getRandomValues()} differs before creating the SafeFrame URL. In this example, each button click resulted in two extra function calls to \texttt{Math.random()} and \texttt{crypto.getRandomValues()}. If the number of function calls to the \texttt{Math.random()} and \texttt{crypto.getRandomValues()} are the same on each replay, then the random values generated will also be the same as in Table \ref{table:SafeFrameSubdomainsReplayTime}. (URL: \url{https://treid003.github.io/random_Values_external_JS_with_async.html} | WACZ: \url{https://zenodo.org/records/13695291/files/2024-09-04-random-values-with-safeframes.wacz?download=1})}
\label{table:differentRandomSafeFrameURLs}
\end{table}

\begin{table}[tbp]
\begin{tabular}{lll}
\multicolumn{3}{c}{Random Values in Google SafeFrame URLs During Replay}                                                                                                                                                                                                                                     \\ \hline
\multicolumn{1}{|c|}{\begin{tabular}[c]{@{}c@{}}Replay Session \\ Number\end{tabular}} & \multicolumn{1}{c|}{\begin{tabular}[c]{@{}c@{}}Generated Using \\ Math.random()\end{tabular}} & \multicolumn{1}{c|}{\begin{tabular}[c]{@{}c@{}}Generated Using \\ crypto.getRandomValues()\end{tabular}} \\ \hline
\multicolumn{1}{|l|}{1}                                                                & \multicolumn{1}{l|}{328a37205b691401e879e520ba757b42}                                         & \multicolumn{1}{l|}{70b75c8cd9262cb991fc0248a75df5d6}                                                    \\ \hline
\multicolumn{1}{|l|}{2}                                                                & \multicolumn{1}{l|}{328a37205b691401e879e520ba757b42}                                         & \multicolumn{1}{l|}{70b75c8cd9262cb991fc0248a75df5d6}                                                    \\ \hline
\multicolumn{1}{|l|}{3}                                                                & \multicolumn{1}{l|}{328a37205b691401e879e520ba757b42}                                         & \multicolumn{1}{l|}{70b75c8cd9262cb991fc0248a75df5d6}                                                    \\ \hline
\multicolumn{1}{|l|}{4}                                                                & \multicolumn{1}{l|}{328a37205b691401e879e520ba757b42}                                         & \multicolumn{1}{l|}{70b75c8cd9262cb991fc0248a75df5d6}                                                    \\ \hline
\multicolumn{1}{|l|}{5}                                                                & \multicolumn{1}{l|}{328a37205b691401e879e520ba757b42}                                         & \multicolumn{1}{l|}{70b75c8cd9262cb991fc0248a75df5d6}                                                    \\ \hline
\multicolumn{1}{|l|}{6}                                                                & \multicolumn{1}{l|}{328a37205b691401e879e520ba757b42}                                         & \multicolumn{1}{l|}{70b75c8cd9262cb991fc0248a75df5d6}                                                    \\ \hline
\multicolumn{1}{|l|}{7}                                                                & \multicolumn{1}{l|}{328a37205b691401e879e520ba757b42}                                         & \multicolumn{1}{l|}{70b75c8cd9262cb991fc0248a75df5d6}                                                    \\ \hline
\multicolumn{1}{|l|}{8}                                                                & \multicolumn{1}{l|}{328a37205b691401e879e520ba757b42}                                         & \multicolumn{1}{l|}{70b75c8cd9262cb991fc0248a75df5d6}                                                    \\ \hline
\multicolumn{1}{|l|}{9}                                                                & \multicolumn{1}{l|}{328a37205b691401e879e520ba757b42}                                         & \multicolumn{1}{l|}{70b75c8cd9262cb991fc0248a75df5d6}                                                    \\ \hline
\multicolumn{1}{|l|}{10}                                                               & \multicolumn{1}{l|}{328a37205b691401e879e520ba757b42}                                         & \multicolumn{1}{l|}{70b75c8cd9262cb991fc0248a75df5d6}                                                    \\ \hline
\end{tabular}
\caption{Random values in Google SafeFrame URLs are the same when replaying the archived web page multiple times with the same number of function calls to the random number functions before creating the Google SafeFrames. (URI-R: \url{https://treid003.github.io/random_Values_external_JS_with_async.html} | WACZ: \url{https://zenodo.org/records/10931734/files/2023-02-21-random-values-with-safeframes.wacz?download=1})} \label{table:SafeFrameSubdomainsReplayTime}
\end{table}

Loading Google ads in a containing web page is an example in which overwriting the seed for the random number generators did not result in the same random numbers being generated during each replay.
In this case, the Google SafeFrame URL's random subdomain differed each time the archived web page was loaded (Table \ref{table:SafeFrameSubdomainChangesWhenLoadingAdDuringReplay}).
The Google SafeFrame subdomain differed in Table \ref{table:SafeFrameSubdomainChangesWhenLoadingAdDuringReplay} in part because a Google SafeFrame was loaded when the ad slot is close to the viewport, not immediately upon replay.  
Delaying the ads' loading can produce different timings in network communications, which ``lead[s] to a varying execution order and thus a different order of pop-requests from the `random' number sequence'' \cite{kiesel-wasp-desires18}.
To explore this phenomenon, we augmented the demo web page (Figure \ref{fig:updatedGSFDemo}) with two sections in which the random values were dynamically generated based on user interaction. The first section featured buttons that, when clicked, generated the random values (Figure \ref{fig:rvButtonClick}). The second section generated random values when the user scrolled to it (Figure \ref{fig:rvScrolling}).
We replayed the updated demo web page five times and changed the number of buttons clicked on during each replay; therefore, the number of function calls to \texttt{Math.random()} and \texttt{crypto.getRandomValues()} differed before reaching the last section. Varying the number of function calls to the random number functions before generating the last two Google SafeFrame URLs produced different random values in the subdomain for the URLs (shown in Table \ref{table:differentRandomSafeFrameURLs}), which is similar to what happened in Table \ref{table:SafeFrameSubdomainChangesWhenLoadingAdDuringReplay}.

If there are multiple JavaScript files making function calls for \texttt{Math.random()} and \texttt{crypto.getRandomValues()} (e.g., running multiple ad services on a web page) before a Google ad is loaded, then the random number sequence will change, which causes variance with the random value included in the Google SafeFrame URL.
In contrast, if the number of function calls to the random number functions is the same, then the random values will be consistent, as in the first version of our demo web page (Table \ref{table:SafeFrameSubdomainsReplayTime}).

Overall, replay systems cannot generate the same random value that was generated during crawl time. Notably, \textit{any} dynamically loaded resources that use random values will encounter this problem. 

\subsubsection{Amazon Ad iframe}
\begin{figure}[tbp]
    \centering
    \includegraphics[width=\textwidth]{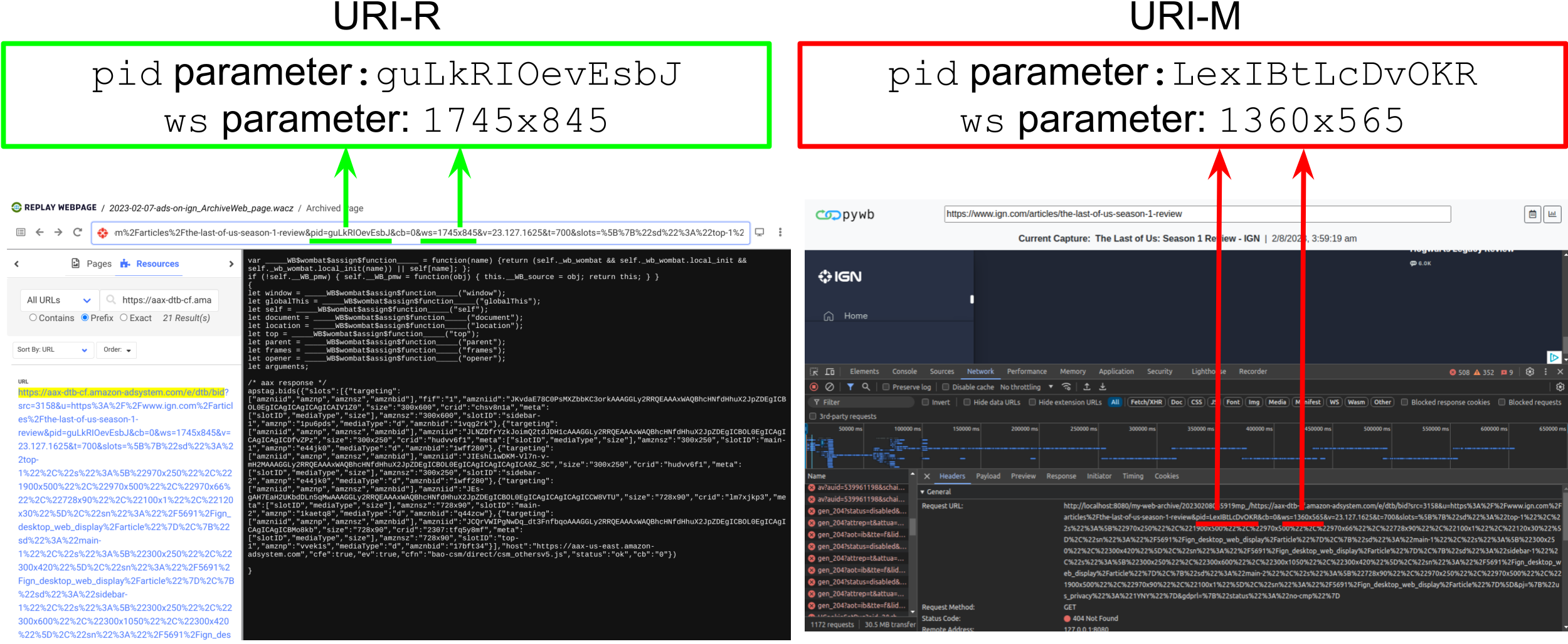}
    \caption{Pywb 2.7.3 failed to load Amazon ads because an ad bid URL failed to load. WACZ: \url{https://zenodo.org/record/8000975/files/2023-02-07-ads-on-ign_ArchiveWeb_page.wacz?download=1} | URI-R: \href{https://aax-dtb-cf.amazon-adsystem.com/e/dtb/bid?src=3158&u=https\%3A\%2F\%2Fwww.ign.com\%2Farticles\%2Fthe-last-of-us-season-1-review&pid=guLkRIOevEsbJ&cb=0&ws=1745x845&v=23.127.1625&t=700&slots=\%5B\%7B\%22sd\%22\%3A\%22top-1\%22\%2C\%22s\%22\%3A\%5B\%22970x250\%22\%2C\%221900x500\%22\%2C\%22970x500\%22\%2C\%22970x66\%22\%2C\%22728x90\%22\%2C\%22100x1\%22\%2C\%22120x30\%22\%5D\%2C\%22sn\%22\%3A\%22\%2F5691\%2Fign_desktop_web_display\%2Farticle\%22\%7D\%2C\%7B\%22sd\%22\%3A\%22main-1\%22\%2C\%22s\%22\%3A\%5B\%22300x250\%22\%2C\%22300x420\%22\%5D\%2C\%22sn\%22\%3A\%22\%2F5691\%2Fign_desktop_web_display\%2Farticle\%22\%7D\%2C\%7B\%22sd\%22\%3A\%22sidebar-1\%22\%2C\%22s\%22\%3A\%5B\%22300x250\%22\%2C\%22300x600\%22\%2C\%22300x1050\%22\%2C\%22300x420\%22\%5D\%2C\%22sn\%22\%3A\%22\%2F5691\%2Fign_desktop_web_display\%2Farticle\%22\%7D\%2C\%7B\%22sd\%22\%3A\%22sidebar-2\%22\%2C\%22s\%22\%3A\%5B\%22300x250\%22\%2C\%22300x600\%22\%2C\%22300x420\%22\%2C\%22100x1\%22\%5D\%2C\%22sn\%22\%3A\%22\%2F5691\%2Fign_desktop_web_display\%2Farticle\%22\%7D\%2C\%7B\%22sd\%22\%3A\%22main-2\%22\%2C\%22s\%22\%3A\%5B\%22728x90\%22\%2C\%22970x250\%22\%2C\%22970x500\%22\%2C\%221900x500\%22\%2C\%22970x90\%22\%2C\%22100x1\%22\%5D\%2C\%22sn\%22\%3A\%22\%2F5691\%2Fign_desktop_web_display\%2Farticle\%22\%7D\%5D&pj=\%7B\%22us_privacy\%22\%3A\%221YNY\%22\%7D&gdprl=\%7B\%22status\%22\%3A\%22no-cmp\%22\%7D}{https://aax-dtb-cf.amazon-adsystem.com/e/dtb/bid?src=3158\&u=https\%3A\%2F\%2Fwww.ign.com\%2Farticles\ ...}}
    \label{fig:differentBidURLs}
\end{figure}

To replay an Amazon ad, a crawler must archive both the embedded web page for the Amazon ad iframe and the ad loaded inside it.
We encountered two challenges with Amazon ads. 
First, although the ad resources were successfully archived, pywb version 2.7.3 \cite{kreymer-pywb-273-gh23} failed to replay some Amazon ads because a URL for the ad bid contained incorrect \texttt{ws} and \texttt{pid} query string parameters (Figure \ref{fig:differentBidURLs}).
The dynamically generated values for these parameters differed during the crawling and replay sessions.
Pywb 2.7.3 only replayed Amazon ads initially loaded inside of a Google SafeFrame. To enable this replay, however, we not only needed to know the URL for the Amazon ad iframe, but also loaded the ad outside of the containing web page.
Second, since the Internet Archive blocked Save Page Now users from archiving the URLs associated with Amazon ad iframes, the Wayback Machine will not replay most Amazon ads.

\begin{figure}[tbp]
    \centering
    \includegraphics[width=\textwidth]{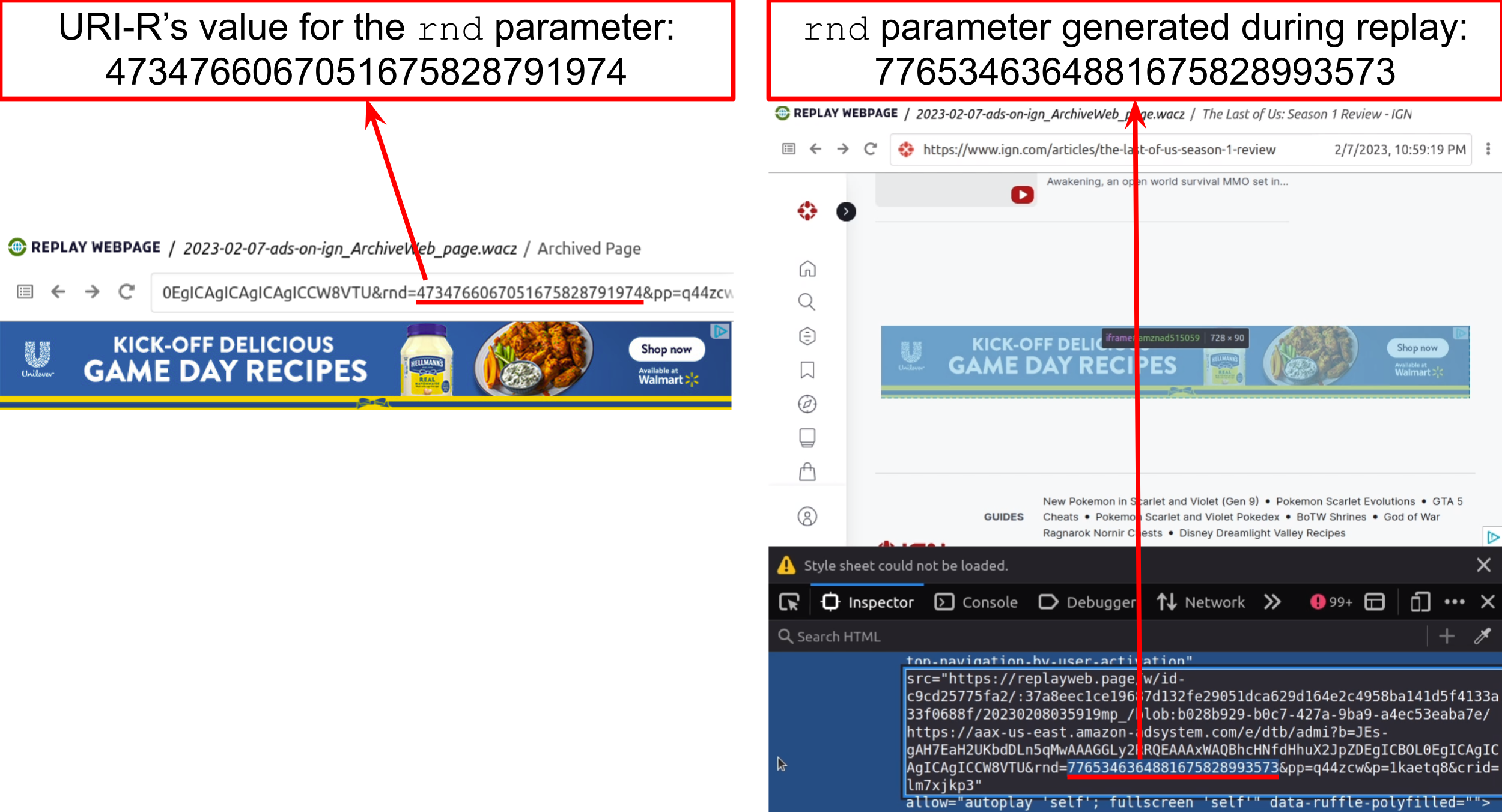}
    \caption{When replaying an Amazon ad iframe, the \texttt{rnd} parameter is not the same as the original value that is in the URI-R. Even though an incorrect URI-M is generated, ReplayWeb.page is able to load the ad. URI-R: \url{https://aax-us-east.amazon-adsystem.com/e/dtb/admi?b=JEs-gAH7EaH2UKbdDLn5qMwAAAGGLy2RRQEAAAxWAQBhcHNfdHhuX2JpZDEgICBOL0EgICAgICAgICAgICCW8VTU&rnd=4734766067051675828791974&pp=q44zcw&p=1kaetq8&crid=lm7xjkp3} | WACZ: \url{https://zenodo.org/record/8000975/files/2023-02-07-ads-on-ign_ArchiveWeb_page.wacz?download=1}}
    \label{fig:differentRndValues}
\end{figure}

Alone among the replay systems we tested, ReplayWeb.page successfully replayed this type of Amazon ad. Amazon ads use a random value in the URL's query string stored in the \texttt{rnd} parameter.
ReplayWeb.page's approach for fuzzy matching made it possible to replay these Amazon ads even when the \texttt{rnd} parameter generated during replay differed from the one generated during crawl time (Figure \ref{fig:differentRndValues}). Requests that have different query string parameters during replay than during crawl time are handled by fuzzy matching to match requests during replay with responses that were captured during crawl time \cite{kiesel-acm18}. 
While the random value generated in the query string of the URL did not prevent ReplayWeb.page from loading an Amazon ad, a random value used in the subdomain of a Google ad's URL did.

Amazon ads' use of random numbers in their iframe URLs caused another problem. Multiple ads may use the same base ad iframe URL, albeit with different query strings. 
This prevents some of the ads from being shown during replay because of how ReplayWeb.page uses fuzzy matching. If multiple ad iframe URLs only differ by their query string, then the same ad will be selected and replayed when loading an unarchived ad iframe URL.

\subsubsection{Loading Embedded Web Page Ads Outside of an Ad iframe} \label{section:flashtalking}
\begin{figure}[tbp]
    \centering
    \includegraphics[width=\textwidth]{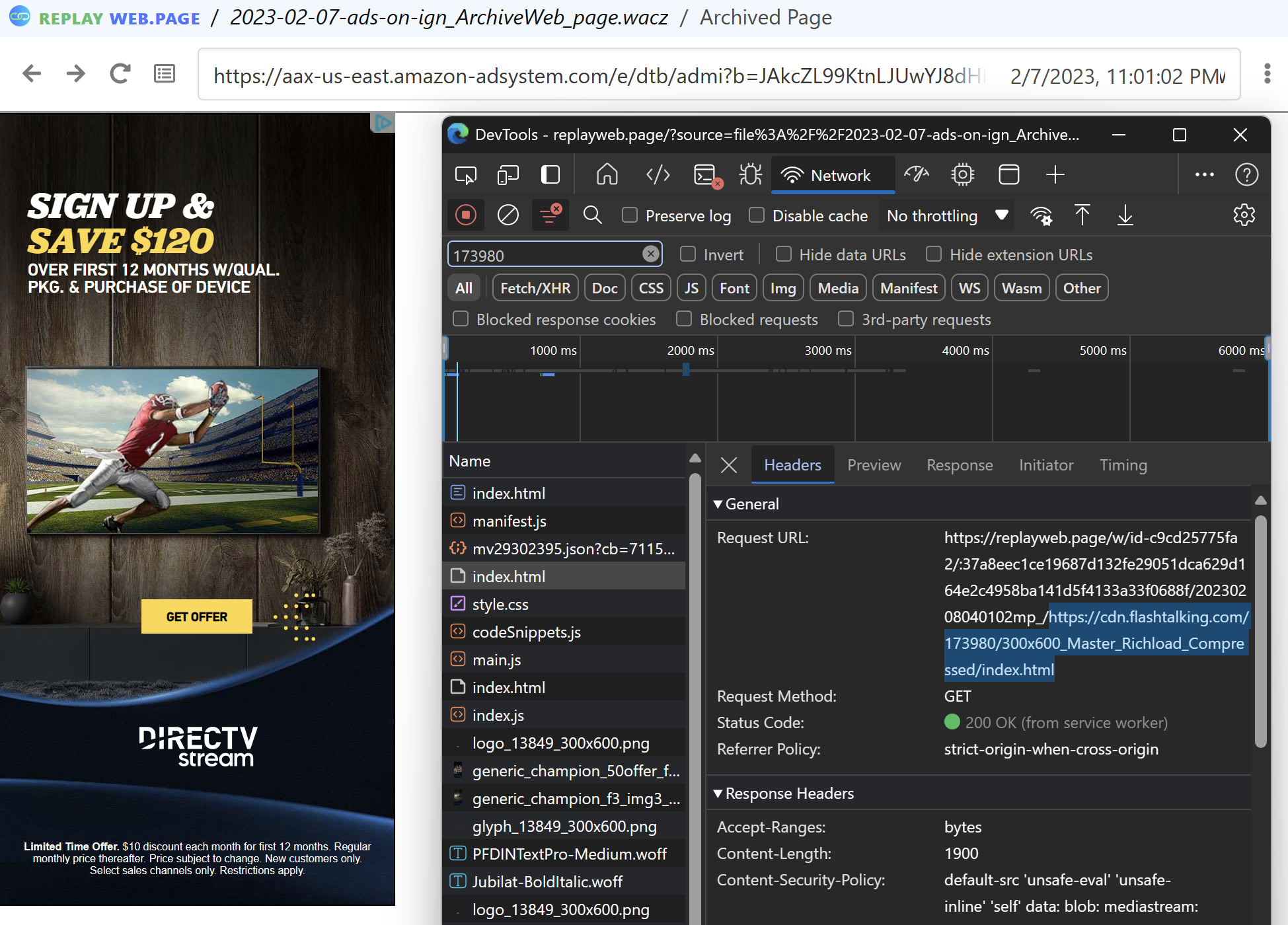}
    \caption{This embedded web page ad will successfully replay when it is loaded inside of an Amazon ad iframe. The correct Richload URI will be loaded when the ad is in the iframe. This ad uses Flashtalking and Amazon ad services. WACZ file: \url{https://zenodo.org/record/8000975/files/2023-02-07-ads-on-ign_ArchiveWeb_page.wacz?download=1} | Ad iframe URI-R: \url{https://aax-us-east.amazon-adsystem.com/e/dtb/admi?b=JAkcZL99KtnLJUwYJ8dHHdIAAAGGLy8dDAEAAAxWAQBhcHNfdHhuX2JpZDEgICBOL0EgICAgICAgICAgICA_KuR0&rnd=8954498773591675828862700&pp=1wff280&p=e44jk0&crid=arcgnw6w} | Richload URI-R: \url{https://cdn.flashtalking.com/173980/300x600_Master_Richload_Compressed/index.html}}
    \label{fig:replayFlashtalkingAdInsideIframe}
\end{figure}

\begin{figure}[tbp]
    \centering
    \includegraphics[width=\textwidth]{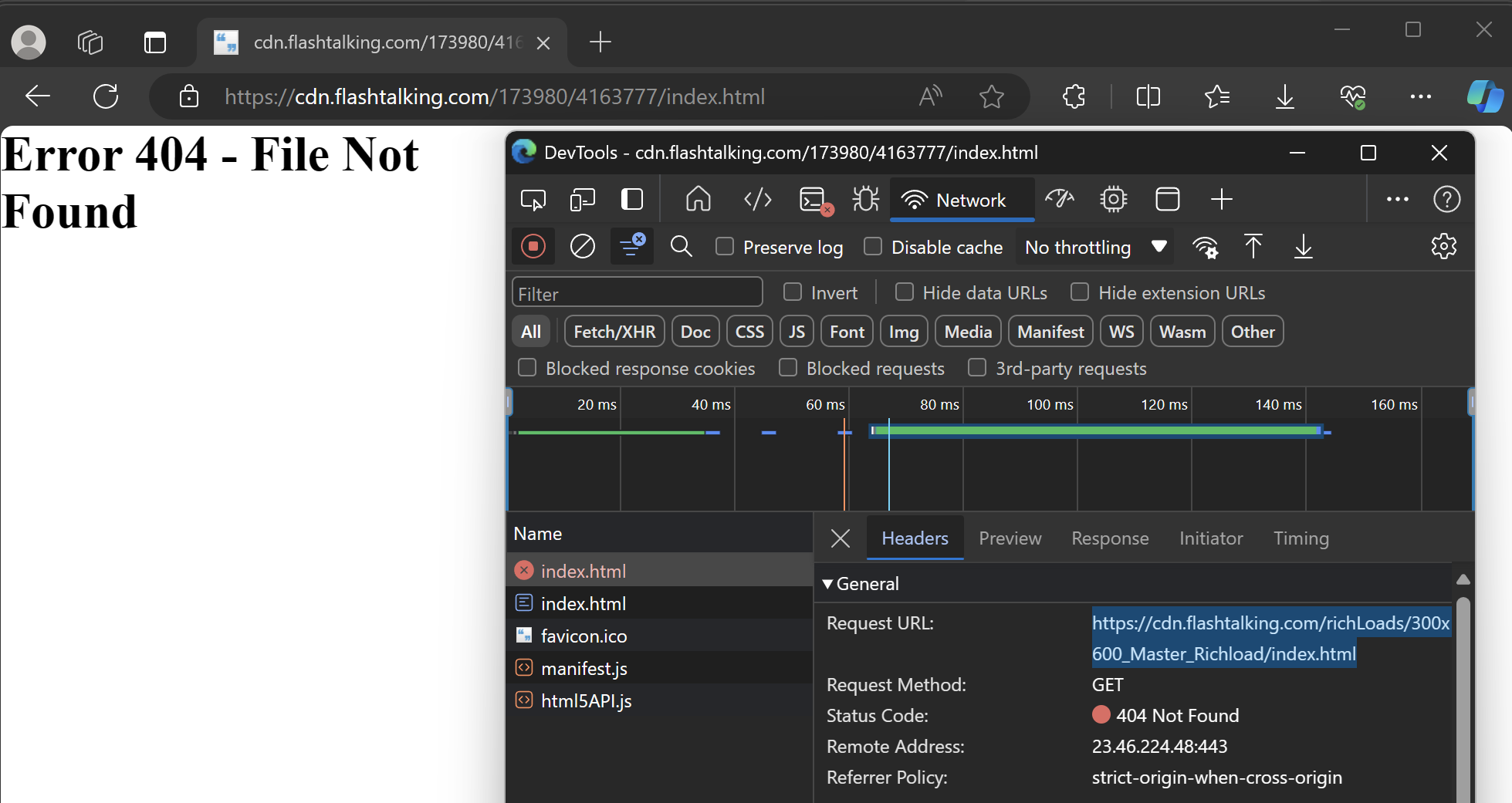}
    \caption{The live web version of this embedded web page ad also fails to load outside of the ad iframe. URI-R: \url{https://cdn.flashtalking.com/173980/4163777/index.html}}
    \label{fig:liveFlashtalkingAdOutsideIframe}
\end{figure}

The JavaScript for Flashtalking’s ad service also dynamically generated a URL that prevented ad replay. This prevented us from replaying the embedded web page ads that we archived during 2023 outside the containing web page because the JavaScript for Flashtalking’s ad service loaded an unarchived web resource. 
\begin{figure}[tbp]
    \centering
    \includegraphics[width=\textwidth]{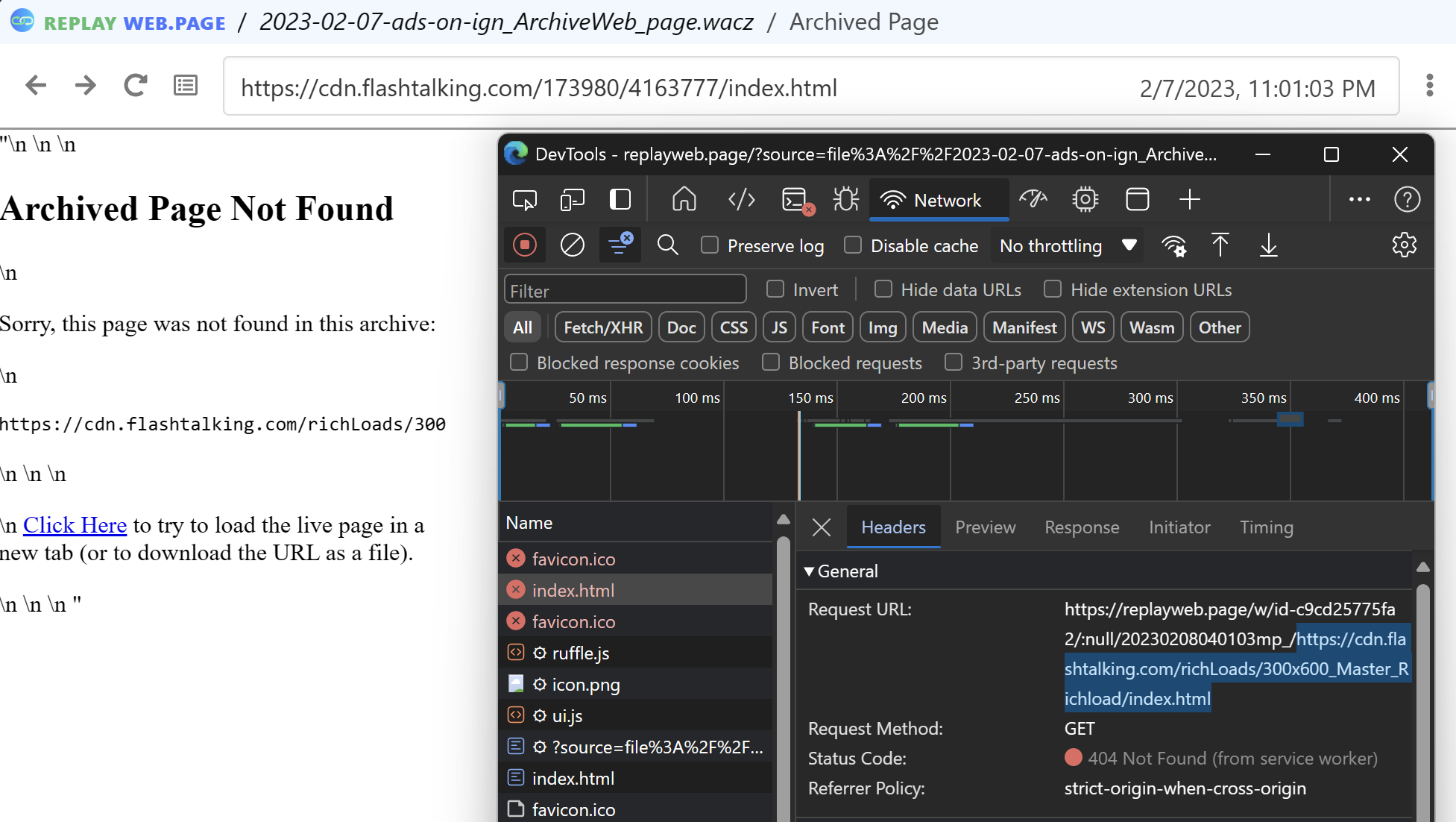}
    \caption{Replaying a successfully archived embedded web page ad outside of its ad iframe. This ad uses Flashtalking and Amazon ad services. WACZ file: \url{https://zenodo.org/record/8000975/files/2023-02-07-ads-on-ign_ArchiveWeb_page.wacz?download=1} | URI-R: \url{https://cdn.flashtalking.com/173980/4163777/index.html}}
    \label{fig:replayFlashtalkingAdOutsideIframe}
\end{figure}
Figure \ref{fig:replayFlashtalkingAdOutsideIframe} shows an example of an embedded web page ad outside of its ad iframe. 
The error message shown in Figure \ref{fig:replayFlashtalkingAdOutsideIframe} is associated with an incorrect resource being loaded that prevents the ad from being replayed. The URI-R \url{https://cdn.flashtalking.com/richLoads/300x600_Master_Richload/index.html} is not associated with the current ad. The correct URI-R that should have been loaded is \url{https://cdn.flashtalking.com/173980/300x600_Master_Richload_Compressed/index.html}, which includes the ad id (173980). The Richload URI includes the ad id when replaying the embedded web page ad in an Amazon ad iframe, which  enables replay of the other ad resources (Figure \ref{fig:replayFlashtalkingAdInsideIframe}).
However, even if we try to access this web page ad outside of the ad iframe on the live web, the web page will use an incorrect Richload URI and the ad will not load~\footnote{Flashtalking might have done this to block ads from loading on a website with which the advertisers do not want to be associated.} (Figure \ref{fig:liveFlashtalkingAdOutsideIframe}). 

\newpage
\subsubsection{Replay of An Ad Can Differ Depending On The Web Browser Used}
\begin{figure}[tbp]
    \centering
    \begin{subfigure}[b]{0.45\textwidth}
        \centering
        \includegraphics[width=\textwidth]{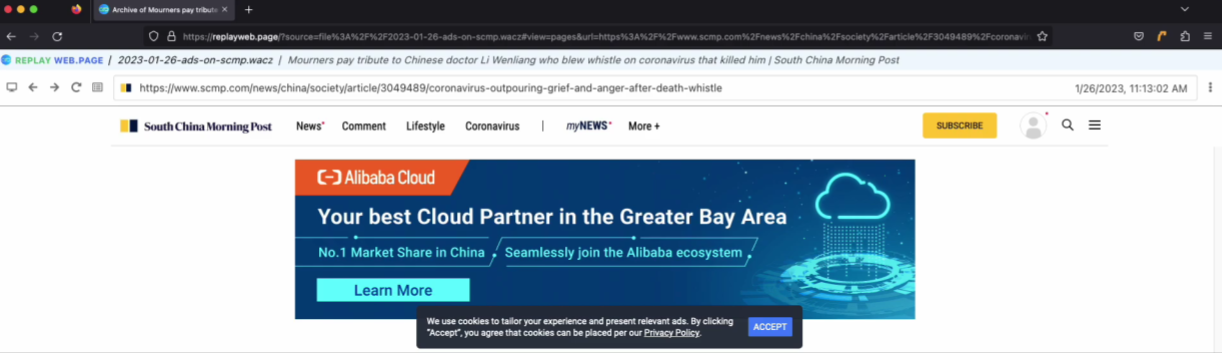} 
        \label{fig:FirefoxReplay}
    \end{subfigure}
    \hfill
    \begin{subfigure}[b]{0.45\textwidth}
        \centering
        \includegraphics[width=\textwidth]{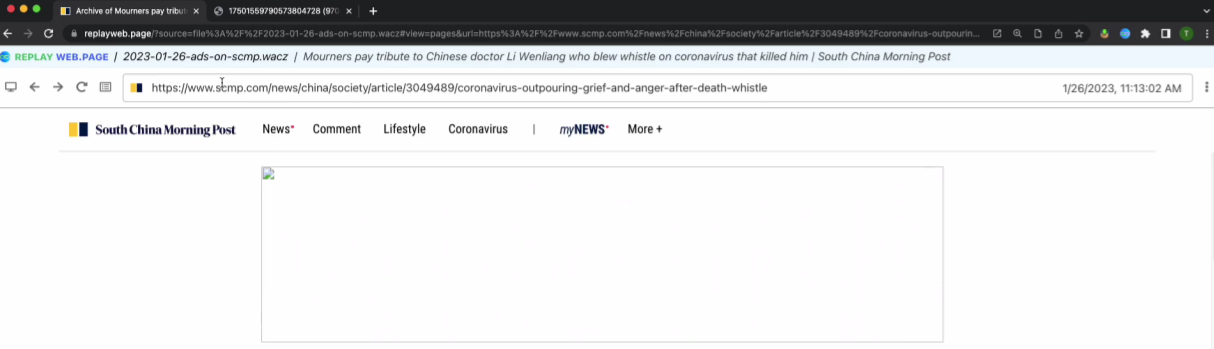}
        \label{fig:ChromeReplay}
    \end{subfigure}
    \caption{The replay of an ad differed depending on the web browser used. WACZ: \url{https://zenodo.org/records/10373131/files/2023-01-26-ads-on-scmp_ArchiveWeb_page.wacz?download=1} | URI-R: \url{https://www.scmp.com/news/china/society/article/3049489/coronavirus-outpouring-grief-and-anger-after-death-whistle} | Video: \url{https://www.youtube.com/watch?v=gCW15i-5teQ}}
    \label{fig:replayDifferentBasedOnBrowser}
\end{figure}

Finally, we identified a replay problem with a replay system (ReplayWeb.page) that used service workers. In January 2023, we archived and replayed a web page~\footnote{\url{https://www.scmp.com/news/china/society/article/3049489/coronavirus-outpouring-grief-and-anger-after-death-whistle}} that included an ad (Figure \ref{fig:replayDifferentBasedOnBrowser}) whose successful replay depended upon the browser used~\footnote{Video: \url{https://youtu.be/gCW15i-5teQ?t=40}}. 
When replay systems use service workers, the replay of an archived web page can differ depending upon a browser's implementation of service workers. We observed this problem when an ad used an iframe with a \texttt{src} attribute value of ``\texttt{about:blank}''.

When we used ReplayWeb.page with Firefox version 109.0 \cite{firefox-109-mozilla23}, the image ad loaded (left side of Figure \ref{fig:replayDifferentBasedOnBrowser}). Conversely, it failed to load when using Chrome version 109.0.5414 \cite{chrome-109} (right side of Figure \ref{fig:replayDifferentBasedOnBrowser}). 
After identifying this problem, we created a GitHub issue \cite{reid-rwp-issue23} on ReplayWeb.page's GitHub repository. One of the comments~\footnote{GitHub issue: \url{https://github.com/webrecorder/replayweb.page/issues/157}} mentioned that the service worker had not gained control of the ad iframe, which led to leaked requests. These leaked requests resulted in a 404 status code during replay for a successfully archived resource. There is a Chromium bug~\footnote{Chromium bug: \url{https://issues.chromium.org/issues/41411856}} related to this issue, where the service worker is unable to access the resources loaded in an ``\texttt{about:blank}'' iframe. 
However, a ReplayWeb.page update fixed this by overriding the \texttt{document.write()} \cite{document-write-mozillat24} method with a blob URL~\footnote{A blob URL is another way to access a File object and it can be used as a \texttt{src} or \texttt{href} attribute \cite{bidelman-11}.} created by the service worker.

\section{Conclusions}
Web advertisements represent a significant and rapidly evolving aspect of digital cultural heritage. The need to preserve them is ever-increasing. But serious problems with archiving and replaying current web ads persist.
This paper explored the process of creating a dataset of 279 recent (January--June, 2023) web ads and discussed the problems we encountered while archiving and replaying these ads.
This dataset was created by archiving 17 web pages from SimilarWeb's top websites worldwide. When archiving these web pages, we utilized four web archiving services (Internet Archive's Save Page Now, Arquivo.pt, archive.today, and Conifer) and three browser-based tools (ArchiveWeb.page, Browsertrix Crawler, and Brozzler). We replayed these archived web pages with four web archiving services (Internet Archive's Wayback Machine, Arquivo.pt, archive.today, and Conifer) and three other replay systems (ReplayWeb.page, pywb, and OpenWayback). 

The process of archiving and replaying these 279 unique web ads yielded five key findings. First, Internet Archive's Save Page Now feature excluded web ads from being archived. Before August 2023, Save Page Now prevented users from archiving ads from well-known ad services like Google AdSense and Amazon Ad Server and it excluded URLs with ad related file and directory names in the URL's path. After August 2023, Save Page Now still excluded ads that were loaded on a web page from being archived, but it did allow an ad's resource(s) to be archived when the user directly archived the URL(s) associated with the ad.
Second, Brozzler was incompatible with recent versions of Google Chrome released after March 2023. This incompatibility prevented Chrome from loading web pages during the crawling session which prevented web resources from being archived.
Third, when executing Google's and Amazon's ad script, the random values generated are not the same during the crawling and replay sessions, because the seed for the random number generator is set by the replay system and the value will not be the same as during crawl time. This resulted in a request for an incorrect URL during replay that was not archived and prevented the ad from loading. 
Fourth, the JavaScript for Flashtalking's ad service prevented the replay of embedded web page ads outside of an ad iframe, because the ad script dynamically generated an incorrect URL that does not exist on the live web. When this incorrect URL is requested, it stops the remaining ad resources from being loaded which resulted in a web page that displayed a 404 error message.
Fifth, some web ads were not loaded during replay depending on the web browser used, because the service worker implementation can differ between browsers. Chromium had a bug that prevented service workers from accessing resources inside an iframe with the \texttt{src} attribute of ``\texttt{about:blank}''. This resulted in leaked requests, which prevented the replay of a successfully archived ad.  
Complementing these findings, we created the Display Archived Ads tool to help find advertisements that were not able to replay when loading the containing web page. Our tool filters out known ad resources that remain invisible during replay, and it displays the live version of an ad alongside the archived version.

Three replay problems need to be fixed.
First, dynamically generated URLs with random values need to be matched with the URLs that were successfully crawled. When performing a URL match, the random value should be removed. Google SafeFrame URLs include the random value in the subdomain and Amazon ad iframe URLs include the random value in the query string. 
Second, when a Flashtalking ad requests a non-existent URL that includes ``\texttt{Richload}'' in the path, it needs to be matched with a different URL that includes ``\texttt{Richload}'' and the ad ID in the URL's path. The ad ID can be retrieved from the URL for the web page ad that initiated the request.
Third, when a replay system uses service workers and is replaying an ad loaded in an iframe with a \texttt{src} attribute value of ``\texttt{about:blank}'', it could replace the \texttt{src} attribute with a blob URL. Using blob URLs for ``\texttt{about:blank}'' iframes is the workaround that ReplayWeb.page used to resolve this problem for Chromium-based browsers.
Fixing these replay problems will not only improve the replay of ads, but also improve the replay of other dynamically loaded embedded resources that use random values or ``\texttt{about:blank}'' iframes.

\section{Acknowledgments}
This research was made possible through the support of the Institute of Museum and Library Services (IMLS) \href{https://www.imls.gov/grants/awarded/lg-252362-ols-22}{\#LG-252362-OLS-22}.

\bibliographystyle{acm}
\bibliography{refs.bib}
\end{document}